\documentclass[preprint,12pt]{elsarticle}

\usepackage[position=above]{caption} 
\usepackage[colorlinks=true,linkcolor=blue,citecolor=blue,urlcolor=blue]{hyperref}
\usepackage[utf8]{inputenc}

\usepackage{amssymb}
\usepackage{amsmath}
\usepackage{pdfpages}
\usepackage{anyfontsize}
\usepackage{orcidlink}
\DeclareMathSizes{12}{10}{8}{6}
\usepackage{makecell, multirow}
\usepackage{algorithm}
\usepackage{algpseudocode}
\usepackage{xcolor} 
\usepackage{subcaption} 
\usepackage{booktabs}
\usepackage{array}
\usepackage{threeparttable}
\usepackage{longtable}
\usepackage{graphicx}
\usepackage{tabularx}

\newlength{\tblwidth}
\begin{document}

\begin{frontmatter}



\title{A Novel Solution for Zero-Day Attack Detection in IDS using Self-Attention and Jensen-Shannon Divergence in WGAN-GP}


\author[inst1]{Ziyu Mu\corref{cor1}\orcidlink{0009-0008-0697-7580}}
\ead{z.mu@lboro.ac.uk}           
\cortext[cor1]{Corresponding author}  

\author[inst1]{Xiyu Shi\orcidlink{0000-0001-6174-3383}}
\ead{x.shi@lboro.ac.uk}

\author[inst1]{Safak Dogan\orcidlink{0000-0002-1465-6495}}
\ead{s.dogan@lboro.ac.uk}

\address[inst1]{Institute for Digital Technologies, Loughborough University, London, UK}

\begin{abstract}
The increasing sophistication of cyber threats, especially zero-day attacks, poses a significant challenge to cybersecurity. Zero-day attacks exploit unknown vulnerabilities, making them difficult to detect and defend against. Existing approaches patch flaws and deploy an Intrusion Detection System (IDS). Using advanced Wasserstein GANs with Gradient Penalty (WGAN-GP), this paper makes a novel proposition to synthesize network traffic that mimics zero-day patterns, enriching data diversity and improving IDS generalization. SA-WGAN-GP is first introduced, which adds a Self-Attention (SA) mechanism to capture long-range cross-feature dependencies by reshaping the feature vector into tokens after dense projections. A JS-WGAN-GP is then proposed, which adds a Jensen-Shannon (JS) divergence-based auxiliary discriminator that is trained with Binary Cross-Entropy (BCE), frozen during updates, and used to regularize the generator for smoother gradients and higher sample quality.  Third, SA-JS-WGAN-GP is created by combining the SA mechanism with JS divergence, thereby enhancing the data generation ability of WGAN-GP. As data augmentation does not equate with true zero-day attack discovery, we emulate zero-day attacks via the leave-one-attack-type-out method on the NSL-KDD dataset for training all GANs and IDS models in the assessment of the effectiveness of the proposed solution. The evaluation results show that integrating SA and JS divergence into WGAN-GP yields superior IDS performance and more effective zero-day risk detection.

\end{abstract}


\end{frontmatter}



\section{Introduction}
\label{sec1}

The cybersecurity landscape changes continually as cyber threats increase in complexity and prevalence. They evolve and propagate during the time elapsed after a software vendor or developer has identified a vulnerability that can be exploited before its public disclosure and the availability of a corresponding patch or remedy. Zero-day attacks exploit these unpatched vulnerabilities in the software or hardware, granting attackers a considerable advantage \cite{10060371}. They can inflict substantial harm, impacting systems from home computers to essential infrastructure \cite{10.1145/3605775}. The consequences of zero-day attacks include unauthorized access to sensitive information, service interruptions, financial losses, and damage to an organization's reputation. For example, the notorious Stuxnet infection exploited several zero-day vulnerabilities in the Windows operating system to cause significant damage to nuclear facilities in Iran\cite{8593143}.

Current approaches to mitigating zero-day attacks are challenging to implement due to the reliance on static patterns or known vulnerabilities \cite{10265713}. Traditional defenses, such as patch management and signature-based Intrusion Detection Systems (IDS), are effective against known threats. However, these methods encounter significant challenges when dealing with new, previously unseen vulnerabilities \cite{9395929}. Meanwhile, attackers continue to develop sophisticated techniques, from polymorphic malware to adversarial code obfuscation, that hinder rapid and accurate detection \cite{10041602, 9272237}. To combat these evolving threats, researchers have explored Machine Learning (ML) and, more recently, advanced generative models such as Generative Adversarial Networks (GANs) \cite{goodfellow2014generativeadversarialnetworks}. In principle, GANs provide a versatile and adaptive framework that can generate realistic samples of complex attack behaviors, even those that have not yet been observed in the past. By exposing IDS models to a broader range of synthetic malicious profiles, defenders can potentially enhance the models' generalization capabilities, making them better at identifying and mitigating zero-day threats. Through this process, GANs can help reduce reliance on established parameters and foster a more proactive security posture.

Consequently, recent studies have employed GAN-based methodologies to address the limitations prevalent in conventional security frameworks. By leveraging the capacity to learn underlying patterns in the attack data, GANs can generate synthetic samples that closely resemble zero-day conditions. Introducing these novel samples into the training process enables IDS models to enhance their ability to detect and respond to previously unknown threats. Therefore, researchers increasingly utilize a variety of GANs to produce diverse malicious samples, enrich scarce training datasets, and improve predictive accuracy in identifying emergent attack vectors.

In an earlier work \cite{mu2024information}, the Wasserstein Generative Adversarial Network with Gradient Penalty (WGAN-GP) model is employed to improve zero-day attack detection. In this paper, three new generative models, termed as Self-Attention-based WGAN-GP (SA-WGAN-GP), WGAN-GP with Jensen-Shannon divergence (JS-WGAN-GP), and WGAN-GP integrated with SA and JS (SA-JS-WGAN-GP), were proposed to synthesize novel attack samples to improve the detection performance of IDS models. Five IDS models implemented with different ML methods, including linear Support Vector Machine (SVM), C4.5 Decision Tree (DT), Deep Neural Network (DNN), Convolutional Neural Network (CNN), and Long Short-Term Memory (LSTM) network, were used to evaluate the effect of the proposed generative models in detecting zero-day attacks. In emulating zero-day attacks using the Leave-One-Attack-Type-Out (LOAO) method on the NSL-KDD dataset, a specific type of attack samples are excluded from training for both the Generator (G) and the IDS. The IDS training dataset also includes synthesized data, and the IDS evaluation is performed on the original test set. This design evaluates whether augmentation enhances generalization to an unseen class, while keeping it distinct from true zero-day discovery.

In summary, our contributions are as follows:
\begin{itemize}

\item SA mechanism and JS divergence in WGAN-GP: Model long-range feature dependencies and promote multimodal coverage, thereby improving the representation of rare and marginal data and reducing GAN model collapse.

\item Dynamic weighted joint loss in the proposed SA-JS-WGAN-GP: Adaptively balances the G, Discriminator (D), and Critic (C), preventing majority class bias and reducing false negatives for atypical patterns.

\item Our experiments on the NSL-KDD dataset \cite{5356528, 425a-3e55-18}, including both binary, multi-classification tasks and LOAO experiment, demonstrate that the three proposed generative models, namely SA-WGAN-GP, JS-WGAN-GP, and SA-JS-WGAN-GP, are able to enhance the generalization ability of IDS models. The generated data proved crucial for IDS training, helping to improve the IDS models' performance in detecting and responding effectively to zero-day attacks.

\end{itemize}

\begin{figure*}[htbp]
\centerline{\includegraphics[width=1.0\textwidth,height=0.35\textwidth]{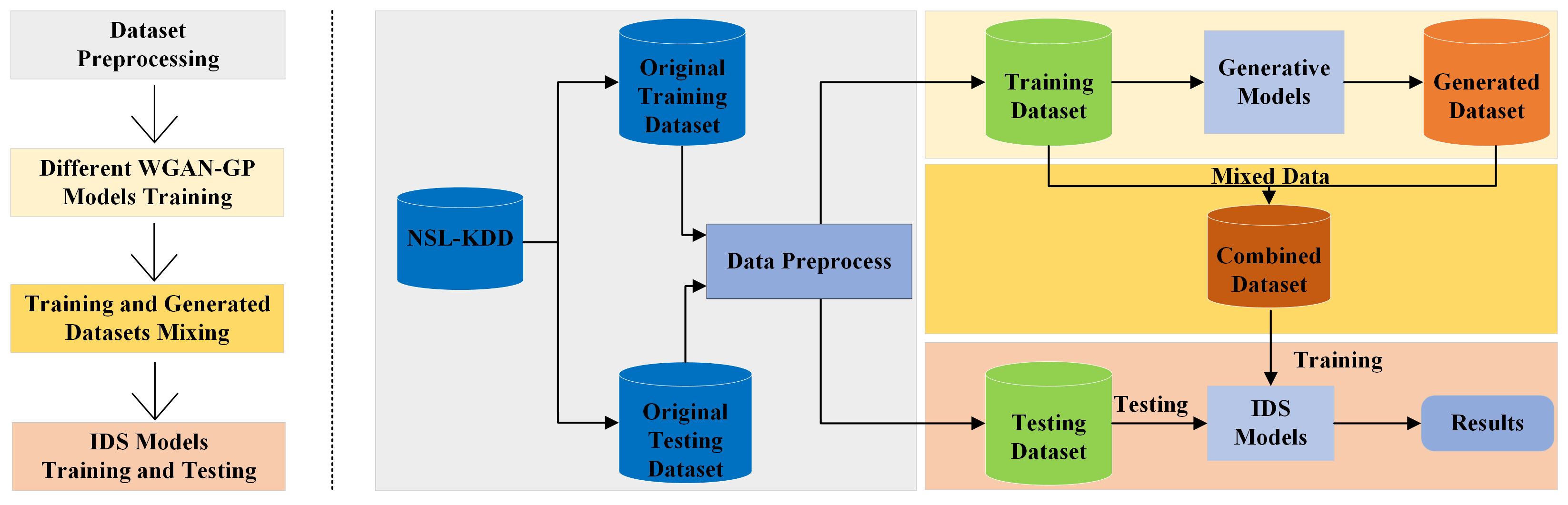}}
\caption{The flowchart (left) shows the workflow of the enhanced performance evaluation process for IDS models, while the right side illustrates each step involved.}
\label{fig.1}
\end{figure*}

Figure \ref{fig.1} illustrates the overall process of employing the proposed SA-, JS-, and SA-JS-WGAN-GP models to enhance the generalization capacity of IDS models in identifying zero-day threats.

This paper is organized as follows: Section II reviews recent studies on detecting zero-day threats using ML, deep learning, and GAN-based models. The research methodologies are described in Section III, including details of the proposed SA-WGAN-GP, JS-WGAN-GP, and SA-JS-WGAN-GP. Section IV explains the experiment setup for evaluation, including the dataset, details of sample generation with the proposed WGAN-GP variants and their hyperparameters, and the IDS models and their parameters. Section V evaluates the performance of the proposed models and discusses the results. Section VI summarizes the study's findings and outlines future research directions.

\section{Related Work}
Researchers have applied ML methods to IDS models to deal with zero-day attacks in network traffic. The original GAN model was employed in \cite{RAHMAN2024101212} to generate different data types for the NSL-KDD database. Eight ML techniques were used for binary classification studies, with Gaussian Naive Bayes attaining a classification accuracy of 84\%. In \cite{fu2021gan}, Fu et al. used Auxiliary Classifier GAN to generate different types of samples for the NSL-KDD database, and their multi-classification test achieved an accuracy of 76\% when using the Multilayer Perceptron model for classification. Zhao et al. introduced the attackGAN model to generate novel attack data on the NSL-KDD dataset \cite{zhao2021attackgan}. Utilizing the generated data, they evaluated the IDS model and got an attack success rate of 81.37\%. This research demonstrates that GANs can produce high-fidelity malicious network traffic that effectively evades black-box IDS models while preserving essential traffic functionality. Their findings underscore critical vulnerabilities in existing IDS frameworks, highlighting the urgent need for more robust and adaptive defense mechanisms.

Nonetheless, owing to intricate network traffic and constrained computational resources, numerous researchers have employed deep learning to manage complex patterns and high-dimensional data. For minority class resampling, an Imbalanced Generative Adversarial Network (IGAN) filter was proposed by \cite{HUANG2020102177}, with convolutional layers being incorporated into the G to enhance its expressive power. This method allowed a 6-layer DNN to achieve an accuracy of 84.45\% in multi-classification evaluations on the NSL-KDD database. Park et al. created a new type of Network Intrusion Detection System (NIDS) that uses Boundary-balanced GAN (BeGAN) and autoencoders to find strange patterns in network traffic data better \cite{9908159}. BeGAN and autoencoders are iteratively trained to improve the ability of anomaly detection in the D, achieve superior data normalization via three data preprocessing techniques (one-hot encoding, outlier analysis, and feature scaling), and augment the model's feature extraction performance on the data. Utilizing a 2-layer CNN model, they achieved accuracies of 90.3\% in binary classification and 93.2\% in multi-classification with the NSL-KDD dataset. 

Zhang and Zhao created a new GAN model with a Variational Autoencoder (VAE) to make high-quality intrusion samples more stable and efficient \cite{9516484}. Several types of data were produced by the generative model from the NSL-KDD dataset and combined with the original training dataset as input into different IDS models, whilst these IDS models are ultimately evaluated with the original test dataset. The experiment showed that in the multi-classification test, a detection accuracy of 81.49\% was recorded after training the CNN-based IDS model with the mixed dataset. In \cite{10530014}, Zhong et al. used a combination of Wasserstein Divergence Objective for GAN (WGAN-div) and an Information Maximizing GAN (InfoGAN) to address the data imbalance issue in intrusion detection. The WGAN-div model was used to oversample the imbalanced data samples in the NSL-KDD dataset to improve the data distribution. The InfoGAN model employed an IDS model for binary and multi-classification assessments, attaining an accuracy of 91.1\% and 90.9\% in the binary and multi-classification evaluations of the NSL-KDD dataset, respectively.

Lim et al. examined how GAN models could alleviate the challenges of anomaly detection caused by data scarcity \cite{LIM2024103733}. Sun et al. proposed an unsupervised dual variational generative adversarial model, termed MTS-DVGAN, using VAE to represent multivariate time series and provide diverse samples \cite{SUN2024103570}. The MTS-DVGAN attained an accuracy of 82.4\% in the multi-classification test of the NSL-KDD dataset. The multi-module DWGF-IDS model, proposed in \cite{GU2023366}, integrates WGAN-GP, SA, and autoencoders to get an 85.1\% multi-classification accuracy on the NSL-KDD dataset.  This methodology enhances the detection rate of unidentified attacks and diminishes false positives for minority attack traffic. Soleymanzadeh and Kashef used a CNN ensemble to create a new GAN architecture, referred to as GANs-ECNN \cite{9850286}. It achieved an accuracy of 86.3\% in multi-classification tests on NSL-KDD. This architecture not only improved IDS models' detection accuracy, it also introduced a more stable training process and accelerated GANs-ECNN model convergence compared to other generative methods. Wan et al. advocated employing a GAN model to sample the NSL-KDD dataset and used the AdaBoost algorithm to integrate multiple DNN classifiers to solve the data imbalance problem \cite{9581127}. This method achieved an accuracy of 82.3\% in multi-classification experiments.

Moreover, owing to the large amounts of data in the NSL-KDD dataset and the extended training time of the GANs, many researchers have optimized the training dataset to ensure that their GAN models do not experience model collapse and unstable training during the training process. Wang et al. proposed a multi-critic WGAN-GP model that generated only 10,000 samples for a limited number of attack categories in NSL-KDD \cite{10577721}. This method achieved an accuracy of 81\% in multi-classification experiments. Combining GAN with a Transformer model, Feng et al. introduced the TransGAN model and got an accuracy of 84.64\% for binary classification\cite{10583331}. In order to reduce the TransGAN's computational cost, they used the XGBoost algorithm \cite{10.1145/2939672.2939785} for dimensionality reduction in the data preprocessing stage, extracting 17 of the most pertinent features from a total of 41 in the NSL-KDD dataset. Recent work emphasized the importance of interpretability in IDS design. For example, the XAI-IDS framework \cite{article3} demonstrates how explainable AI can improve transparency and robustness in cybersecurity, complementing generative and detection models.

In general, previous GAN-based IDS research had primarily focused on enhancing supervised classifiers using WGAN-GP, CGAN, VAEGAN, and TransGAN variants employing adversarial unsupervised pipelines. Most studies utilized known category testing, mixing generated samples back into the training set, and reporting gains on the official testing set. While the NSL-KDD testing set includes a few attacks not included in the training set, explicit zero-day attack samples are rarely used. This leaves two gaps: limited coverage of minority and marginal data, and weak modeling of long-range feature dependencies. Research on both of these gaps would aid model generalization under unseen attack variants.

\section{Proposed Methodology}

\subsection{From WGAN to WGAN-GP}

To solve the problems of training instability and model collapse in GAN models, the Wasserstein GAN (WGAN) was introduced  \cite{arjovsky2017wassersteingan, Creswell_2018}, which used the Wasserstein distance as a loss function. The WGAN model is characterized by replacing the Critic (C) with a function and requiring it to be 1-Lipschitz continuity. C is used to train the G by providing a score for generated samples. The loss for C is based on the Wasserstein distance between the real and generated distributions, penalized by the gradient of C's score with respect to interpolated samples. The Lipschitz constraint limits the function's maximum gradient to ensure the Wasserstein distance's validity. The C and G of the WGAN model use different loss functions \(L_{C}\) and \(L_{G}\), respectively, defined as:

\begin{equation}
L_{C}  = \mathbb{E}_{z\sim{P}_{\text{z}}(z)}[C(G({z}))] -  \mathbb{E}_{x\sim P_{\text{data}}(x)}[C({x})],
\end{equation}
\begin{equation}
L_{G} = -\mathbb{E}_{z\sim{P}_{\text{z}}(z)}[C(G({z}))],
\end{equation} where \(\mathbb{E} \) represents the expectation value, which helps average the performance of C and G on the distribution of the real and generated data; \(z \) is random noise; \( P_{\text{data}}(x) \) is the real sample distribution; \(P_{z}(z) \) is the latent noise distribution; \( C(x) \) represents the probability that \( x \) is a real sample; \(G(z) \) is the generated sample.

To address the limitations of weight clipping, Gulrajani et al. \cite{gulrajani2017improvedtrainingwassersteingans} introduced a Gradient Penalty (\(GP \))  based on the WGAN model to enforce the Lipschitz constraint more effectively. The \(GP \) encourages the gradient norm to be close to 1, enforcing the Lipschitz constraint without directly limiting the weights. The \(GP \) term is calculated as \(GP \) = \( \lambda \mathbb{E}_{\hat{x} \mathbb{P}_{\hat{x}}}[(\|\nabla_{\hat{x}} C(\hat{x})\|_{2} - 1)^{2}] \), where $\lambda$ is the gradient norm used to enforce the Lipschitz constraint; $\hat{x}$ is obtained by taking the weighted average of \( P_{\text{data}}(x) \) and \(P_{z}(z) \); $\nabla$ denotes the gradient operator used to calculate the gradient of the C function concerning its input; and \(\|\nabla_{\hat{x}} C(\hat{x})\|_{2} \) is the L2 norm of the gradient. This modification significantly alleviates issues associated with unstable training and model collapse. The formulas for calculating $\hat{x}$ and the loss functions of C and G in the WGAN-GP model are as follows, with \( \epsilon \) as a crucial parameter that provides control over how much influence the original data versus the generated data has in forming the new data point:

\begin{equation} \hat{x} = \epsilon x + (1 - \epsilon)G(z), \epsilon \in (0, 1), \end{equation} \begin{equation} L_{C} = \mathbb{E}_{{z} \sim {P}_{z}}[C({G(z)})] - \mathbb{E}_{x \sim {P}_{data}}[C(x)] + GP, \end{equation} \begin{equation} L_{G} = -\mathbb{E}_{{z} \sim {P}_{z}}[C({G(z)})]. \end{equation}

\subsection{The SA-WGAN-GP Model}

In Natural Language Processing (NLP), the SA mechanism extends the Transformer architecture to capture complex dependencies between data in a sequence \cite{zhang2019selfattentiongenerativeadversarialnetworks}. The integration of the SA mechanism with WGAN-GP represents a novel contribution to the design of generative models for intrusion detection. This approach aligns with broader efforts to leverage machine learning techniques in enhancing IDS, as discussed in \cite{article}. Its primary goal is to capture long-range dependencies and relationships within the data, making it an indispensable component in various deep learning applications, such as language translation software used in everyday life. Unlike traditional models that rely on local operations, such as convolution, the SA mechanism highlights the importance and relevance of different parts of the input sequence to one another, irrespective of their positions in the sequence \cite{shaw2018selfattentionrelativepositionrepresentations}.

The fundamental concept of the SA mechanism is to compute a weighted sum of values, with the similarity between the query and a collection of keys dictating the weight (attention score) \cite{9093190}. Each element in the input data (whether it is a word in a sentence or a pixel in an image) is converted into three vectors: a Query vector (\(Q\)), a Key vector (\(K\)), and a Value vector (\(V\)). The attention score is determined by the scaled dot product of the query and key vectors, followed by a softmax operation to normalize the sum to 1.

Given an input sequence \( X = [x_1, x_2, \dots, x_n] \), each input element \( x_i \), \(i \in [1, n]\), is linearly transformed into Query (\(Q_i\)), Key (\(K_i\)), and Value (\(V_i\)) vectors, as follows:
    \begin{equation}
    Q_i = W_Q x_i, \quad K_i = W_K x_i, \quad V_i = W_V x_i, 
    \end{equation}where \( W_Q, W_K, W_V \) are learned weight matrices.

The attention score for each pair of input elements \( (Q_i, K_j) \), \(i, j \in [1, n]\),  is computed as the dot product between \( Q_i \) and \( K_j \), scaled by the square root of \( d_k \) - the dimension of the Key vectors \(K\) to stabilize training and control gradient magnitudes, as shown in (\ref{formula.7}).
    \begin{equation}
    \text{Attention Score}_{(Q_i,K_j)} = \frac{Q_i \cdot K_j^T}{\sqrt{d_k}}
     \label{formula.7}
    \end{equation}

Then, these scores are processed through a softmax function to yield the attention weights  \( \alpha_{ij} \), as defined in (\ref{formula.8}), ensuring they are positive and sum to one.
    \begin{equation}
    \alpha_{ij} = \text{Softmax}\left(\frac{Q_i \cdot K_j^T}{\sqrt{d_k}}\right) = \frac{\exp\left(\frac{Q_i \cdot K_j^T}{\sqrt{d_k}}\right)}{\sum_{j=1}^{n} \exp\left(\frac{Q_i \cdot K_j^T}{\sqrt{d_k}}\right)}
    \label{formula.8}
    \end{equation}

The output for each input element \( x_i \) is computed as a weighted sum of the \(V\) in (\ref{formula.9}), where the weights are the attention weights \( \alpha_{ij} \).
    \begin{equation}
    \text{Output}_i = \sum_{j=1}^{n} \alpha_{ij} V_j
    \label{formula.9}
    \end{equation}

The set of weighted sums for all input elements gives the final output of the SA for the entire sequence, as shown in (\ref{formula.10}).
    \begin{equation}
    \text{Self-Attention}(X) = [\text{Output}_1, \text{Output}_2, \dots, \text{Output}_n]
    \label{formula.10}
    \end{equation}

The SA mechanism allows deep learning models to incorporate all input data. The model's ability to represent long-range dependencies, combined with the feasibility of parallel computing, makes it an essential component of contemporary neural network architectures, particularly those used for generative tasks like those executed by GANs.

Consequently, a SA-WGAN-GP model is created by incorporating the SA mechanism into the WGAN-GP framework. The SA-WGAN-GP model analyzes the intricate, high-dimensional data within the NSL-KDD dataset and extracts the complicated interdependencies across features. In traditional models, including basic GANs and WGAN-GP, it is difficult to effectively capture these long-range dependencies, resulting in impractical data generation and even model collapse. The SA mechanism addresses this issue by enabling the model to dynamically focus on different parts of the input data to capture global dependencies between features. This characteristic enables the SA-WGAN-GP model to generate more nuanced and diverse data, reflecting the data distribution features in the NSL-KDD dataset.

\subsection{The JS-WGAN-GP Model}

A JS-divergence-based D was incorporated into the WGAN-GP framework, resulting in the JS-WGAN-GP model consisting of a G, a D, and a C. The D is trained with a Binary Cross-Entropy (BCE) loss to approximate the JS divergence between real and generated distributions, and its gradients are passed only to G as a regularizer without updating C. This modification aims to enhance the quality and diversity of the data generated during training on the NSL-KDD dataset. While WGAN-GP effectively mitigates model collapse and stabilizes training by utilizing the Wasserstein distance, it still faces challenges in fully distinguishing between complex, overlapping distributions inherent in the data \cite{10.1145/3460418.3479301}. Integrating the JS divergence alongside the Wasserstein distance provides an additional measure that sharpens the distinction between distributions that are similar but not identical \cite{borji2018prosconsganevaluation}. By complementing the Wasserstein distance, the JS divergence allows for a more discriminative evaluation of generated samples, thus enhancing the diversity and fidelity of the synthetic data. Consequently, the JS-WGAN-GP model is capable of generating synthetic data that represents more accurately the nuanced patterns present in real network traffic, ultimately improving the generalization ability of the IDS in detecting zero-day attacks.

JS divergence is a symmetrized and smooth variant of Kullback-Leibler (KL) divergence, offering a more refined similarity measure between two probability distributions \cite{MENENDEZ1997307}. The enhanced capacity to assess distributional similarity enables the JS-WGAN-GP model to produce data that accurately reflects the underlying distribution, especially in intricate datasets such as NSL-KDD.

Let \( P_G(x) \) denote the distribution of generated samples induced by $G(z)$ with $z \sim P_z$. Given two probability distributions \( P_{\text{data}}(x) \) and \( P_G(x) \) , the JS divergence \( D_{JS}(P_{\text{data}} \| P_G) \) is defined as:

\begin{equation} D_{JS}(P_{\text{data}} \| P_G) = \frac{1}{2} D_{KL}(P_{\text{data}} \| M) + \frac{1}{2} D_{KL}(P_G \| M), \end{equation} where \( D_{KL}(\cdot \| \cdot) \) represents the KL divergence, \( M = \frac{1}{2}(P_{\text{data}} + P_G) \) is the mixed distribution, which is the average of the real and generated distributions.

The KL divergence between two distributions \(P\) and \(Q\) is given by:

\begin{equation} D_{KL}(P \| Q) = \int_{\mathbf{x}} P(\mathbf{x}) \log\left(\frac{P(\mathbf{x})}{Q(\mathbf{x})}\right) d\mathbf{x}. \end{equation}

This measures the relative entropy between two distributions, quantifying how one distribution diverges from another.

Unlike the KL divergence, the JS divergence is symmetric:

\begin{equation} D_{JS}(P_{\text{data}} \| P_G) = D_{JS}(P_G \| P_{\text{data}}). \end{equation}

It is also bounded by

\begin{equation} 0 \leq D_{JS}(P_{\text{data}} \| P_G) \leq \log(2). \end{equation}

This feature makes JS divergence a more stable and reliable indicator of distribution similarity, especially when the distributions are almost identical.

Although the JS divergence is defined between the distributions of real data and the generated data in the proposed model, it is approximated by training a JS-based D with BCE, whose gradient feedback serves as a JS-based regularizer for G. The JS-WGAN-GP model combines the Wasserstein distance and JS divergence with a dynamically adjusted weight \( \lambda_{JS} \) to balance their contributions. The loss function \( L_{JS} \) of JS-based D, and the regularization  \(L_G^{\mathrm{JS}}\) to G in JS-WGAN-GP are defined as:
\begin{equation}
L_{JS}
= \mathbb{E}_{x \sim P_{\text{data}}}\!\Big[-\log \sigma(D_{\text{JS}}(x))\Big]
+ \mathbb{E}_{z \sim P_z}\!\Big[-\log\!\big(1 - \sigma(D_{\text{JS}}(G(z)))\big)\Big],
\end{equation}
\begin{equation}
L_G^{\mathrm{JS}} = \mathbb{E}_{z \sim P_z}\!\left[-\log \sigma(D_{\text{JS}}(G(z)))\right],
\end{equation}
where $\sigma(\cdot)$ denotes the sigmoid function
\begin{equation}
\sigma(u) = \frac{1}{1 + e^{-u}}.
\end{equation} The overall adversarial loss of JS-WGAN-GP is expressed as $L_{\text{C}}^{\mathrm{JS}} = L_C + L_{JS}$. The total loss function of G in JS-WGAN-GP denotes 
\begin{equation}
L_G^{\mathrm{total}} = L_G + \lambda_{JS}\,L_G^{\mathrm{JS}}.
\end{equation}

To reduce tuning overhead while keeping the Wasserstein comparable with the JS regularization, \(\lambda_{JS}\) is updated conservatively using the loss ratio \(r=L_C/L_{JS}\) in every 10 epochs as follows:
\begin{equation}
\lambda_{JS} \;\leftarrow\;
\begin{cases}
1.05\,\lambda_{JS}, & r > 1, \\[4pt]
0.95\,\lambda_{JS}, & r \leq 1 ,
\end{cases}
\label{formula.19}
\end{equation}
and is clipped such that \(\lambda_{JS}\in[0.1,\,10]\). Given that each generative experiment runs for 10,000 epochs per model, we observed a $\pm$5\% step for \(\lambda_{JS}\) offers adequate granularity for stable adaptation without overreacting to short-term noise.

During the JS-WGAN-GP training process, supplementary input can improve G's learning and motivate it to represent the data distribution range accurately. This model integrates Wasserstein distance and JS divergence to equilibrate the global context and improve training stability. It aims to strengthen WGAN-GP's stability and efficacy by addressing issues such as model collapse and gradient disappearance, utilizing JS divergence to improve the quality of generated data. The model's capacity to generate data enhances the generalization performance of the IDS model, thereby improving its detection of zero-day attacks.

\subsection{The SA-JS-WGAN-GP Model}

\begin{figure*}[htbp]
\centerline{\includegraphics[width=0.9\textwidth,height=0.5\textwidth]{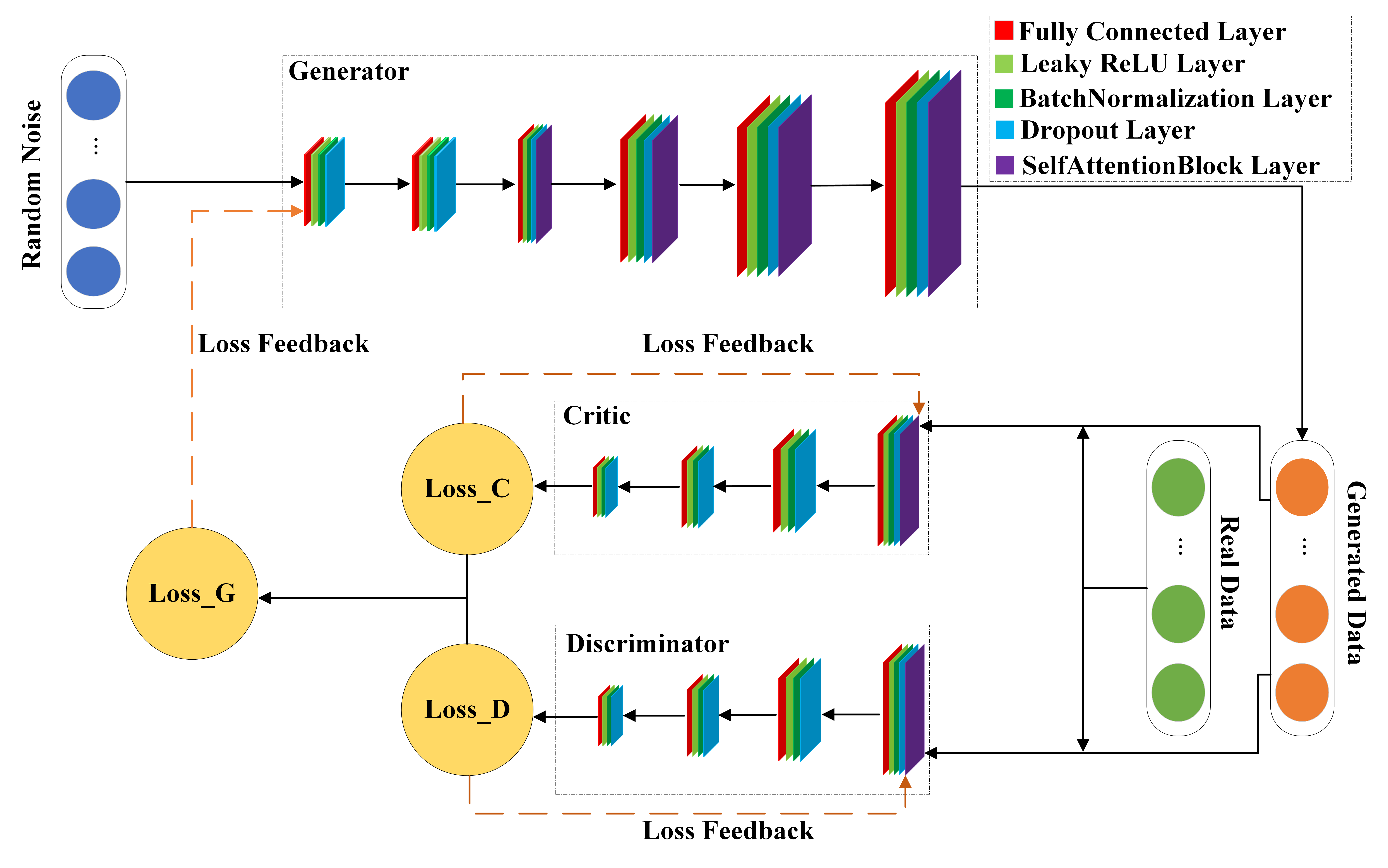}}
\caption{The structure of the proposed SA-JS-WGAN-GP. Loss\_G, Loss\_D, and Loss\_C represent the loss values of G, D, and C, respectively.}
\label{fig.2}
\end{figure*}

Extending the SA-WGAN-GP and JS-WGAN-GP, the SA-JS-WGAN-GP model was developed to combine JS divergence with Wasserstein loss to enhance gradient smoothness and leverage the SA mechanism to capture global dependencies. It is envisaged that data generated by the SA-JS-WGAN-GP can significantly enhance the IDS models' generalization, improving their effectiveness in detecting zero-day attacks. In the SA-JS-WGAN-GP model, two distinct components are introduced:

Figure \ref{fig.2} illustrates the network architecture of our proposed SA-JS-WGAN-GP model. \textbf{Algorithm \ref{alg:sa_js_wgan_gp}} presents the pseudocode used for training the model.

\begin{algorithm}[!htbp]
\scriptsize
\caption{Pseudo Code for Generating Data for SA-JS-WGAN-GP}
\label{alg:sa_js_wgan_gp}
\begin{algorithmic}[1]
\Require Dataset $\mathcal{X}\subset\mathbb{R}^{d}$, batch size $m$, critic steps $n_{\text{critic}}$, Adam LRs $(\alpha_C,\alpha_G,\alpha_D)$, weights $(\lambda_{\mathrm{GP}},\lambda_{\mathrm{JS}})$, epochs $E$
\Ensure Trained generator $G$, Wasserstein critic $C$, and JS discriminator $D$
\Statex \textbf{Notation.} $G(\mathbf{z};\theta_G):\mathbb{R}^{k}\!\to\!\mathbb{R}^{d}$, $C(\mathbf{x};\theta_C):\mathbb{R}^{d}\!\to\!\mathbb{R}$, $D(\mathbf{x};\theta_D):\mathbb{R}^{d}\!\to\!\mathbb{R}$ (logits). Noise $\mathbf{z}\!\sim\!p(z)$; real $\mathbf{x}^{(r)}\!\sim\!\mathcal{X}$; fake $\mathbf{x}^{(g)}\!=\!G(\mathbf{z})$. Interpolation $\hat{\mathbf{x}}=\epsilon \mathbf{x}^{(r)}+(1-\epsilon)\mathbf{x}^{(g)}$, $\epsilon\!\sim\!\mathcal{U}(0,1)$.
\Statex \textbf{Tabular SA (in $G$ and $C$).} Insert self-attention blocks after dense projections:\; Dense $\rightarrow$ Reshape $(L,1)$ $\rightarrow$ SelfAttention $+$ residual $\rightarrow$ BN/activation $\rightarrow$ Flatten.
\Statex \textbf{BCE\_logits.} $\operatorname{BCE\_logits}(u,y)=-\big[y\log\sigma(u)+(1-y)\log(1-\sigma(u))\big]$, $\sigma(u)=1/(1+e^{-u})$.
\State Initialize $\theta_G,\theta_C,\theta_D$; Adam optimizers with $(\alpha, \beta_1{=}0.5,\beta_2{=}0.9)$.
\For{epoch $=1$ to $E$}
  \State\textbf{(A) Critic update: $n_{\text{critic}}$ times; update $\theta_C$ only}
  \For{$t=1$ to $n_{\text{critic}}$}
    \State Sample $\{\mathbf{x}^{(r)}_i\}_{i=1}^m \!\sim\! \mathcal{X}$,\; $\{\mathbf{z}_i\}_{i=1}^m\!\sim\!p(z)$;\; set $\mathbf{x}^{(g)}_i=G(\mathbf{z}_i;\theta_G)$
    \State Scores $s^{(r)}_i=C(\mathbf{x}^{(r)}_i;\theta_C)$,\; $s^{(g)}_i=C(\mathbf{x}^{(g)}_i;\theta_C)$
    \State Interpolate $\hat{\mathbf{x}}_i=\epsilon_i \mathbf{x}^{(r)}_i+(1-\epsilon_i)\mathbf{x}^{(g)}_i$, $\epsilon_i\!\sim\!\mathcal{U}(0,1)$
    \State Gradient penalty $\mathrm{GP}_i=\big(\|\nabla_{\hat{\mathbf{x}}_i} C(\hat{\mathbf{x}}_i)\|_2-1\big)^2$
    \State \textbf{Critic loss:}\; $L_C=\frac{1}{m}\sum_i s^{(g)}_i - \frac{1}{m}\sum_i s^{(r)}_i + \lambda_{\mathrm{GP}}\cdot\frac{1}{m}\sum_i \mathrm{GP}_i$
    \State \textbf{Update:}\; $\theta_C \leftarrow \theta_C - \alpha_C \nabla_{\theta_C} L_C$ \Comment{\emph{Only $C$ is updated here}}
  \EndFor
  \Statex
  \State\textbf{(B) JS discriminator update: once; update $\theta_D$ only}
  \State Sample $\{\mathbf{x}^{(r)}_i\}_{i=1}^m$, $\{\mathbf{z}_i\}_{i=1}^m$;\; $\mathbf{x}^{(g)}_i=G(\mathbf{z}_i;\theta_G)$
  \State Logits $u^{(r)}_i=D(\mathbf{x}^{(r)}_i;\theta_D)$,\quad $u^{(g)}_i=D(\mathbf{x}^{(g)}_i;\theta_D)$
  \State \textbf{JS/BCE loss:}\; $L_D=\frac{1}{m}\sum_i \Big[\operatorname{BCE\_logits}(u^{(r)}_i,1)+\operatorname{BCE\_logits}(u^{(g)}_i,0)\Big]$
  \State \textbf{Update:}\; $\theta_D \leftarrow \theta_D - \alpha_D \nabla_{\theta_D} L_D$ \Comment{\emph{$G$/$C$ frozen; no updates}}
  \Statex
  \State\textbf{(C) Generator update: once; update $\theta_G$ only}
  \State Sample $\{\mathbf{z}_i\}_{i=1}^m$;\; $\mathbf{x}^{(g)}_i=G(\mathbf{z}_i;\theta_G)$
  \State \textbf{Wasserstein term:}\; $L_G^{\mathrm{W}} = -\frac{1}{m}\sum_i C(\mathbf{x}^{(g)}_i;\theta_C)$ \Comment{\emph{$C$ used for scores; $C$ frozen}}
  \State \textbf{JS regularizer:}\; $L_G^{\mathrm{JS}} = \frac{1}{m}\sum_i \operatorname{BCE\_logits}\!\big(D(\mathbf{x}^{(g)}_i;\theta_D),1\big)$ \Comment{\emph{$D$ provides logits; $D$ frozen}}
  \State \textbf{Total:}\; $L_G = L_G^{\mathrm{W}} + \lambda_{\mathrm{JS}}\, L_G^{\mathrm{JS}}$
  \State \textbf{Update:}\; $\theta_G \leftarrow \theta_G - \alpha_G \nabla_{\theta_G} L_G$ \Comment{\emph{Only $G$ is updated}}
  \Statex
  \State \textbf{(Optional) $\lambda_{\mathrm{JS}}$ schedule.} For all reported results, keep $\lambda_{\mathrm{JS}}$ constant. Optionally, every $k$ epochs: compute ratio $r=\frac{|L_C|}{L_D+\varepsilon}$; if $r>1$ set $\lambda_{\mathrm{JS}}\leftarrow\min(10,1.05\lambda_{\mathrm{JS}})$ else $\lambda_{\mathrm{JS}}\leftarrow\max(0.1,0.95\lambda_{\mathrm{JS}})$.
\EndFor
\State \textbf{Output.} Return $G$ (generator), $C$ (critic), $D$ (JS discriminator)
\end{algorithmic}
\end{algorithm}

\section{Experimental Setup}

\subsection{Dataset For the Proposed WGAN-GP Models}

The NSL-KDD dataset is widely acknowledged and utilized in the development and assessment of IDS models, offering a dependable and precise foundation for training and evaluating algorithms for IDS models \cite{8988230}. The dataset contains more than 120,000 entries and is balanced between normal (53\%) and abnormal (47\%) records, as described in \cite{8925239}. Each entry comprises 41 attributes, including 3 nominal, 6 binary, and 32 numeric attributes. The KDDTrain and KDDTest subsets of the NSL-KDD dataset are employed for training and testing the proposed IDS models. To validate the generating capacity of the three proposed WGAN-GP models, the data preprocessing methodology outlined in \cite{mu2024information} is employed. During the generative training of the JS-WGAN-GP and SA-JS-WGAN-GP models, a constraint on the preprocessed data is imposed, limiting the data range to [-1.0, 1.0]. The attacks in the NSL-KDD dataset can be classified broadly into four categories:

\begin{itemize}
\item Denial of Service (DoS): An attack designed to render a resource unavailable, typically by overwhelming the target with excessive Internet traffic.
\item Probe: Involves scanning for open ports and identifying exploitable vulnerabilities.
\item User-to-Root (U2R): An attack in which the perpetrator gains root access after initially compromising a normal user account.
\item Remote-to-Local (R2L): This involves unauthorized access and exploitation of local user privileges and differs from U2R in that the attacker initially has no account on the target system.
\end{itemize}

However, the NSL-KDD dataset is a legacy benchmark and does not fully reflect modern encrypted or application-layer network traffic, nor today’s attack mix. It is used here as a controlled, widely replicated tabular benchmark to compare generative augmentation methods and implement a reproducible zero-day evaluation based on the standard train and test split. Similar considerations are discussed in \cite{Zaki_Naser_2024}, which outlines both the advantages and limitations of NSL-KDD for zero-day intrusion detection. To address its limitations on zero-day attacks, a LOAO experiment is carried out, in which the R2L attacks are removed from all training data for all the proposed SA-, JS-, and SA-JS-WGAN-GPs and IDS models, and the IDS performance is measured on that unseen type of attacks to assess whether synthetic augmentation improves zero-day attack detection.

\subsection{IDS Models}

In the experiment, five different ML models are used in the IDS:
\begin{itemize}
\item A `linear' SVM model was implemented for outlier detection \cite{Cristianini_Shawe-Taylor_2000}.
\item The C4.5 DT algorithm \cite{10.5555/583200} was also implemented and used. C4.5 can handle datasets with both categorical and numerical attributes efficiently. It utilizes a thresholding system for continuous attributes to split the data, making it versatile for different data types \cite{8024318}.
\item In addition, a two-layer DNN model was used in the experiment. The first layer is composed of 32 neurons, and the second layer comprises 16 neurons.
\item Furthermore, a 1D CNN model with two convolutional layers was experimented with. The convolutional layer consists of 64 filters with a kernel size of 5, followed by a fully connected layer with 16 neurons. In addition, a Max pooling layer with a window size of 3 is added to the first convolutional layer. A batch normalization layer is added after each convolutional layer. ReLU is used for all activation functions.
\item Finally, a two-layer LSTM model was adopted for the experiment. Both layers have 64 LSTM cells, and a fully connected layer with 32 neurons is added after the second layer.
\end{itemize}

The DT, DNN, CNN, and LSTM-based models were trained 50 times with random seeds ranging from 42 to 91 to reflect variation from weight initialization and optimization. The linear SVM-based IDS model is deterministic for a given dataset and hyperparameter setting, so multiple seeds are not needed.

In all experiments, the number of IDS training epochs was set to 100. Despite the stability of these model architectures, each IDS model was independently trained fifty times.

The IDS models were developed using Tensor-Flow v2.8, CUDA v11.3, and Cudnn v8.2 and trained on a Linux workstation with Ubuntu 20.04.5 LTS OS, NVIDIA GeForce RTX3090Ti GPU with 24GB of  VRAM, and Intel Core i7-12700KF CPU.

\subsection{Synthesizing Dataset for IDS}

We employed WGAN-GP and the three proposed SA-, JS-, and SA-JS-WGAN-GP models to generate various data types from the NSL-KDD dataset in training different IDS models. The five significant data types in NSL-KDD, namely Normal, R2L, Probe, DoS, and U2R, were used to train the different WGAN-GP models. Due to the data imbalance in NSL-KDD, which can hinder the models' ability to learn diverse data characteristics, synthetic data were generated in proportion to the original dataset distribution. Specifically, 50,000 Normal data points were generated, along with 20,000 each for DOS, Probe, and R2L, and 10,000 for U2R.

In the generative experiments, SA-, JS-, and SA-JS-WGAN-GP utilized a batch size of 256 and Adam optimizer with learning rate lr=$10^{-4}$, $\beta_1$=0.5, $\beta_2$=0.9, and gradient clipping is 1.0. All weights were initialized using the He-uniform method. Training was performed for 10,000 epochs without early termination.  Regularization was implemented via Batch Normalization, Dropout, and the Gradient Penalty in WGAN-GP with \(\lambda_{GP} = 10\). All features were numerical and normalized to the range [-1.0, 1.0], with categorical variables encoded during preprocessing. The coefficient $\lambda_{\text{JS}}$ is initialized to 1.0 and updated according to ({\ref{formula.19}}), which was incorporated into the JS-WGAN-GP model. The generating process was continuously monitored by documenting the classification of each synthetic sample as either normal or an attack.

In both binary and multi-classification experiments, the IDS models were trained on a combined dataset that merges the generated data of the proposed WGAN-GP models with the original training dataset, KDDTrain. Consequently, there are four different combined datasets, corresponding to the basic WGAN-GP and three proposed WGAN-GP variants. The data categories in the combined training dataset were reduced to normal and abnormal in binary classification, with DoS, Probe, R2L, and U2R all being categorized as abnormal. In the multi-classification experiment, the combined training dataset included the original dataset and generated data for Probe, R2L, and U2R, addressing the data imbalance issue in the original NSL-KDD dataset. This strategy prevents the dominant category from overwhelming the model and enhances the model's ability to accurately distinguish between different types of attacks.

\begin{figure*}[htbp]
    \centering
    \scriptsize
    \begin{subfigure}{0.45\textwidth}
        \centering
        \includegraphics[width=\textwidth]{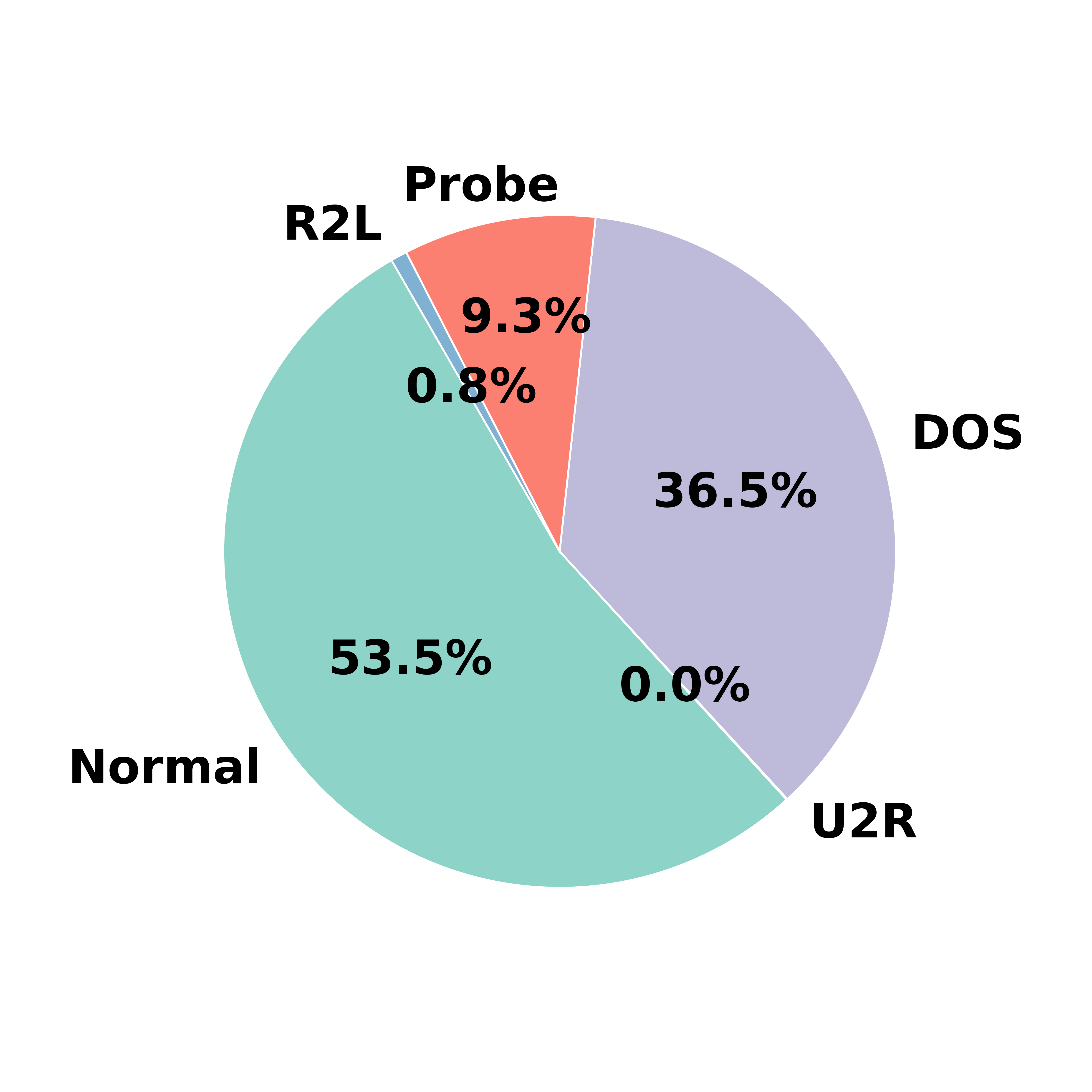}
        \caption{}\vspace{-0.5em}  
        \label{fig:subfig_a}
    \end{subfigure}
    \hfill
    \begin{subfigure}{0.45\textwidth}
        \centering
        \includegraphics[width=\textwidth]{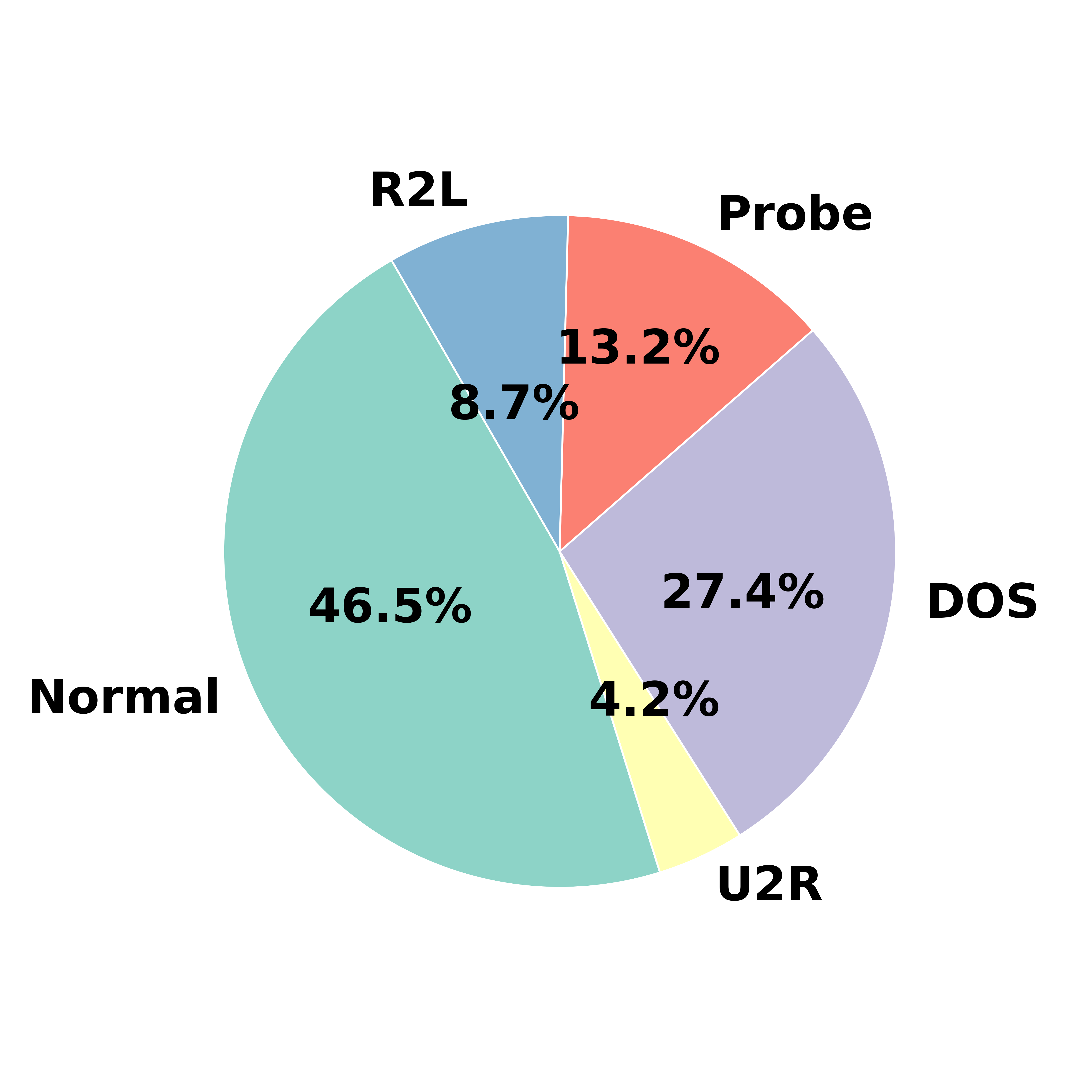}
        \caption{}\vspace{-0.5em}  
        \label{fig:subfig_b}
    \end{subfigure}
    \vspace{0.2\baselineskip}  

    \begin{subfigure}{0.45\textwidth}
        \centering
        \includegraphics[width=\textwidth]{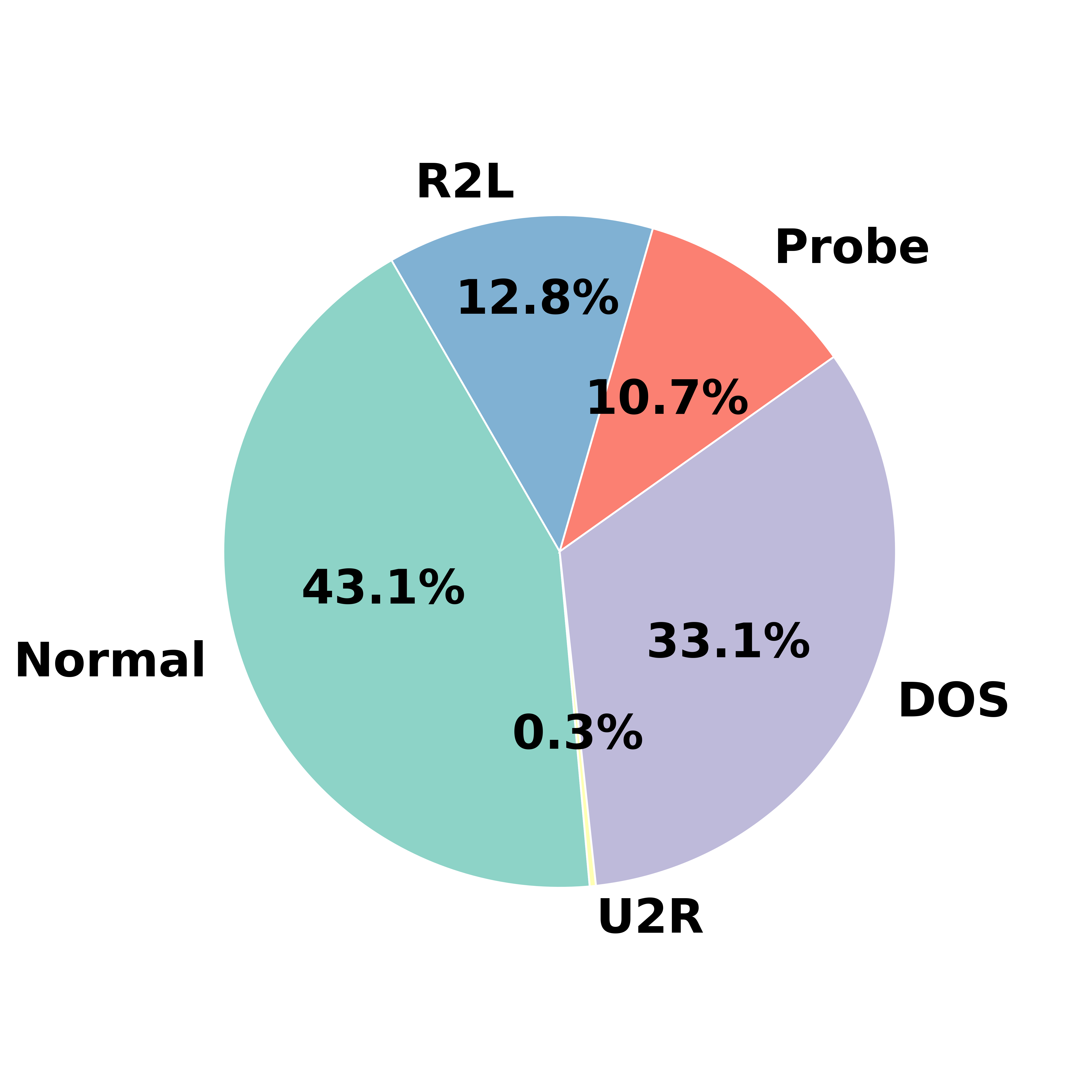}
        \caption{}\vspace{-0.5em}
        \label{fig:subfig_c}
    \end{subfigure}
    \hfill
    \begin{subfigure}{0.45\textwidth}
        \centering
        \includegraphics[width=\textwidth]{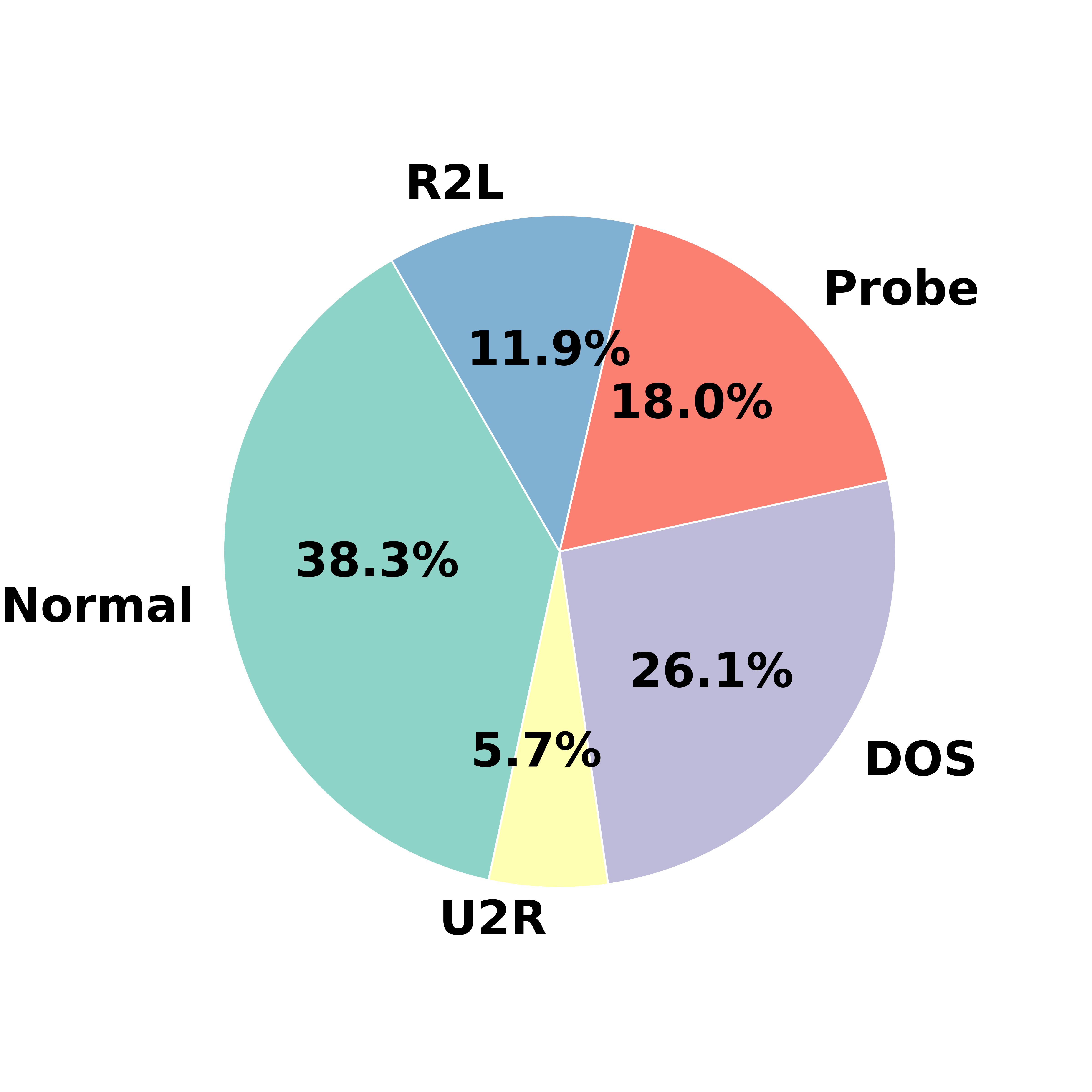}
        \caption{}\vspace{-0.5em}
        \label{fig:subfig_d}
    \end{subfigure}

    \caption{Distribution of data types of the training and testing dataset in the NSL-KDD dataset. 
    (a) The original training dataset contains a significant proportion of Normal (53.46\%) and DOS (36.46\%) data, while other attack types, like R2L (0.8\%) and U2R (0.04\%), are rare. 
    (b) Synthetic data has been generated and added to the binary classification training dataset, balancing the attack types. U2R now has a higher proportion (4.18\%) compared to the original dataset. 
    (c) The original NSL-KDD test dataset has proportions similar to the original training set. 
    (d) Synthetic data has been generated and added to the multi-class training set, showing a more balanced distribution of attack types, likely helping the model better recognize each category.}
    \label{fig.3}
\end{figure*}

Figure \ref{fig.3} shows the distribution of various data categories in the original and experimental datasets before and after merging the generated data on the NSL-KDD dataset. Table \ref{tab:wgan_gp_architectures} shows the structures and hyperparameters adopted by the various proposed WGAN-GP models.

\begin{table}[!ht]
    \centering
    \scriptsize

    \begin{threeparttable} [t]
        \caption{Architectures and Training Parameters of Various WGAN-GP Models}
        \label{tab:wgan_gp_architectures}

      \begin{tabularx}{\textwidth}{%
          >{\raggedright\arraybackslash}X  
          >{\raggedright\arraybackslash}X           
          >{\raggedright\arraybackslash}X           
          >{\raggedright\arraybackslash}X           
          >{\raggedright\arraybackslash}X           
      }
      \toprule
      & \textbf{WGAN-GP} & \textbf{SA-WGAN-GP} & \textbf{JS-WGAN-GP} & \textbf{SA-JS-WGAN-GP} \\
      \midrule
      \textbf{G} 
          & Dense layers: 128, 256, 512
          & Dense layers: 128, 256, 512 + SA layers
          & Dense layers: 64, 64, 128, 256, 512, 1024
          & Dense layers: 64, 64, 128, 256, 512, 1024, Dense layers in SA: 128, 256, 512 \\
      \addlinespace[2ex]
      \textbf{C} 
          & Dense layers: 512, 256, 128
          & Dense layers: 512, 256, 128 + SA layers
          & Dense layers: 128, 64, 32, 16, Output
          & Dense layers: 128, 64, 32, 16, Output, Dense layers in SA: 128 and 64 \\
      \addlinespace[2ex]
      \textbf{D} 
          & --
          & --
          & Dense layers: 128, 64, 32, 16, Output
          & Dense layers: 128, 64, 32, 16, Output, Dense layers in SA: 128 and 64 \\
      \addlinespace[2ex]
      \textbf{Learning Rate} 
          & G: $1\times10^{-4}$, C: $1\times10^{-4}$
          & G: $1\times10^{-4}$, C: $1\times10^{-4}$
          & G: $1\times10^{-4}$, C: $1\times10^{-4}$, D: $1\times10^{-4}$
          & G: $1\times10^{-4}$, C: $1\times10^{-4}$, D: $1\times10^{-4}$ \\
      \addlinespace[2ex]
      \textbf{Epochs}
          & 10,000
          & 10,000
          & 10,000
          & 10,000 \\
      \addlinespace[2ex]
      \textbf{G Activation Function} 
          & LeakyReLU, Tanh
          & LeakyReLU, Tanh
          & LeakyReLU, Tanh
          & LeakyReLU, Tanh \\
      \addlinespace[2ex]
      \textbf{C Activation Function}
          & LeakyReLU
          & LeakyReLU
          & LeakyReLU
          & LeakyReLU \\
      \addlinespace[2ex]
      \textbf{D Activation Function} 
          & --
          & --
          & LeakyReLU, Sigmoid
          & LeakyReLU, Sigmoid \\
      \bottomrule
      \end{tabularx}

    \end{threeparttable}

\end{table}

\section{Results And Discussion}

This section evaluates the three proposed WGAN-GP variants against the original WGAN-GP baseline. An IDS model is first trained on KDDTrain combined with synthetic samples and validated on the standard KDDTest split. Comparative experiments then measure how different training sets affect IDS performance in both binary and multiclass settings, followed by a focused analysis of results. Lastly, an LOAO experiment is conducted to assess whether synthetic augmentation improves zero-day attack detection. The proposed SA-JS-WGAN-GP is also benchmarked against GAN models reported in the literature on NSL-KDD. Further analysis of the experimental results is given in detail using k-nearest neighbors (kNN) and maximum mean discrepancy (MMD) to analyze the generated data. Limitations are summarized at the end. Except for those with specific notes, all results are presented as mean values in percentage (\%) with standard deviations across 50 trials.

\subsection{Evaluation Metrics}

Four metrics are utilized to evaluate the performance of IDS models, namely the detection accuracy, precision, recall, and F1-score. True Positives (TP) is the count of instances where the model correctly classifies the positive class. True Negatives (TN) is the count of instances where the model correctly classifies the negative class. False Positives (FP) is the count of instances where the model incorrectly classifies a negative class as positive. False Negatives (FN) is the count of instances where the model wrongly classifies the positive class as negative. The four metrics can be calculated as in (\ref{formula.accuracy}), (\ref{formula.precision}), (\ref{formula.recall}), and (\ref{formula.f1}).

\begin{equation}
\text{Accuracy} = \frac{\text{TP} + \text{TN}}{\text{TP} + \text{FP} + \text{FN} + \text{TN}}
\label{formula.accuracy}
\end{equation}

\begin{equation}
\text{Precision} = \frac{\text{TP}}{\text{TP} + \text{FP}}
\label{formula.precision}
\end{equation}

\begin{equation}
\text{Recall} = \frac{\text{TP}}{\text{TP} + \text{FN}}
\label{formula.recall}
\end{equation}

\begin{equation}
\text{F1-score} = 2 \times \frac{\text{Precision} \times \text{Recall}}{\text{Precision} + \text{Recall}}
\label{formula.f1}
\end{equation}

For the LOAO evaluation, IDS performance is assessed using the Area Under the Receiver Operating Characteristic curve (AUROC) and the True Positive Rate at a fixed False Positive Rate of 5\% (TPR@5\%FPR).

AUROC measures the probability that a randomly chosen positive instance is ranked higher than a randomly chosen negative instance. It is computed as the area under the Receiver Operating Characteristic (ROC) curve, which plots the True Positive Rate (TPR) against the False Positive Rate (FPR) across all classification thresholds. Formally, the rates are defined as:

\begin{equation}
\text{TPR} = \frac{TP}{TP + FN}, \qquad 
\text{FPR} = \frac{FP}{FP + TN}.
\end{equation}
The AUROC is then calculated by integrating TPR over the range of FPR:
\begin{equation}
\text{AUROC} = \int_{0}^{1} \text{TPR}(\text{FPR}) \, d(\text{FPR}).
\end{equation}

TPR@5\%FPR evaluates detection power under strict conditions by fixing the false positive rate at 5\% and recording the corresponding TPR value from the ROC curve. A higher AUROC score indicates stronger overall discriminative ability across thresholds, while a higher TPR@5\%FPR reflects greater sensitivity to true attacks under low false alarm rates, both of which are desirable in zero-day attack detection.

\subsection{Binary Classification}

\begin{table}[!ht]
    \centering
    \scriptsize  
    \setlength{\tabcolsep}{3pt}
    \renewcommand{\arraystretch}{0.95}
    \begin{threeparttable}
    \caption{Binary Classification Comparison of Different IDS Models for Various WGAN-GP Variants}
    \label{tab.2}
    \begin{tabular}{llc ccc ccc}
        \toprule
        \multirow{2}{*}{WGAN Variant} & \multirow{2}{*}{IDS Model} & \multirow{2}{*}{Acc.} & \multicolumn{3}{c}{Normal} & \multicolumn{3}{c}{Abnormal} \\
        \cmidrule(lr){4-6} \cmidrule(lr){7-9}
         &  &  & Rec. & Prec. & F1-score & Rec. & Prec. & F1-score \\
        \midrule
        \multirow{5}{2cm}{\textbf{WGAN-GP}}
         & SVM  & 75.1$\pm$0.0 & 88.1$\pm$0.0  & 65.7$\pm$0.0  & 75.3$\pm$0.0  & 65.2$\pm$0.0  & 87.9$\pm$0.0  & 74.9$\pm$0.0 \\
         & DT   & 78.8$\pm$1.0  & 91.1$\pm$1.8  & 70.4$\pm$2.2  & 80.9$\pm$1.3  & 49.5$\pm$3.5  & 95.3$\pm$1.6  & 80.3$\pm$2.1 \\
         & DNN  & 75.7$\pm$6.0  & 76.5$\pm$9.1  & 83.1$\pm$11.4  & 78.4$\pm$2.0  & 76.5$\pm$9.1  & 83.1$\pm$11.4  & 78.4$\pm$2.0 \\
         & CNN  & 76.6$\pm$5.1 & 62.7$\pm$10.6  & 96.5$\pm$2.0  & 75.3$\pm$7.5  & 96.7$\pm$2.4  & 66.8$\pm$5.9  & 78.8$\pm$3.4 \\
         & LSTM & \textbf{81.4}$\pm$0.7  & 97.4$\pm$0.2  & 72.7$\pm$2.4  & 82.1$\pm$0.7  & 72.4$\pm$3.4  & 96.9$\pm$0.5  & 82.8$\pm$2.1 \\ 
        \addlinespace[1.5ex]
        \multirow{5}{2cm}{\textbf{SA-WGAN-GP}}
         & SVM  &73.3$\pm$0.0 & 93.6$\pm$0.0  & 62.7$\pm$0.0  & 73.3$\pm$0.0  & 57.9$\pm$0.0  & 92.2$\pm$0.0  & 71.2$\pm$0.0 \\
         & DT   & 78.3$\pm$0.8  & 91.7$\pm$0.8  & 68.5$\pm$1.0  & 78.5$\pm$0.7  & 68.2$\pm$1.5  & 91.6$\pm$0.2  & 78.2$\pm$1.0 \\
         & DNN  & 78.5$\pm$1.5 & 97.2$\pm$0.2  & 67.4$\pm$1.5  & 79.6$\pm$1.0  & 64.3$\pm$2.4  & 96.9$\pm$0.2  & 77.3$\pm$1.8 \\
         & CNN  & \textbf{80.9}$\pm$1.3  & 95.9$\pm$0.9  & 70.5$\pm$1.4  & 81.2$\pm$0.9  & 69.6$\pm$2.1  & 95.7$\pm$0.8  & 80.5$\pm$1.4 \\
         & LSTM & \textbf{80.9}$\pm$1.2 & 96.5$\pm$1.2  & 70.0$\pm$1.3  & 81.5$\pm$0.8  & 69.6$\pm$2.1  & 95.7$\pm$0.8  & 80.5$\pm$1.4 \\ 
        \addlinespace[1.5ex]
        \multirow{5}{2cm}{\textbf{JS-WGAN-GP}}
         & SVM  & 73.1$\pm$0.0 & 98.2$\pm$0.0  & 61.8$\pm$0.0  & 73.1$\pm$0.0  & 54.2$\pm$0.0  & 97.5$\pm$0.0  & 69.6$\pm$0.0 \\ 
         & DT   & 81.3$\pm$0.8  & 94.7$\pm$0.4  & 73.6$\pm$0.9  & 82.9$\pm$0.6  & 74.3$\pm$1.1  & 94.9$\pm$0.4  & 83.4$\pm$0.7 \\
         & DNN  & \textbf{82.3}$\pm$1.6  & 92.1$\pm$4.9  & 73.8$\pm$3.4  & 81.8$\pm$1.1  & 74.9$\pm$5.5  & 93.0$\pm$3.4  & 82.7$\pm$2.2 \\ 
         & CNN  & 81.8$\pm$2.0  & 96.6$\pm$0.5  & 71.5$\pm$2.4  & 82.2$\pm$1.5  & 70.8$\pm$3.5  & 96.5$\pm$0.4  & 81.6$\pm$2.3 \\ 
         & LSTM & 80.6$\pm$1.7  & 95.4$\pm$1.8  & 70.4$\pm$2.2  & 80.9$\pm$1.3  & 69.5$\pm$3.5  & 95.3$\pm$1.6  & 80.3$\pm$2.1 \\
        \addlinespace[1.5ex]
        \multirow{5}{2cm}{\textbf{SA-JS-WGAN-GP}}
         & SVM  & 73.3$\pm$0.0 & 98.2$\pm$0.0  & 62.0$\pm$0.0  & 73.3$\pm$0.0  & 54.5$\pm$0.0 & 97.5$\pm$0.0  & 69.9$\pm$0.0 \\
         & DT   & \textbf{83.1}$\pm$0.6  & 90.0$\pm$1.1  & 77.0$\pm$0.5  & 83.0$\pm$0.4  & 79.7$\pm$0.7  & 91.4$\pm$0.8  & 85.1$\pm$0.3 \\
         & DNN  & 81.6$\pm$3.0 & 89.8$\pm$9.7  & 73.7$\pm$3.1  & 80.6$\pm$4.0  & 75.4$\pm$5.7  & 91.6$\pm$6.7  & 82.3$\pm$2.5 \\
         & CNN  & 81.8$\pm$1.4 & 95.8$\pm$1.7  & 71.6$\pm$1.6  & 81.9$\pm$1.3  & 71.2$\pm$2.1  & 95.8$\pm$1.7  & 81.6$\pm$1.6 \\
         & LSTM & 82.0$\pm$3.3 & 96.2$\pm$1.8  & 71.9$\pm$3.7  & 82.2$\pm$2.6  & 71.3$\pm$5.1  & 96.1$\pm$1.9  & 81.8$\pm$3.6 \\
        \bottomrule
    \end{tabular}
    \begin{tablenotes}\footnotesize
        \item \textbf{Note:} The best average accuracy of different IDS models under different WGAN-GP variants is highlighted in bold.

    \end{tablenotes}
    \end{threeparttable}

\end{table}

\begin{table}[t]
    \centering
    \scriptsize  
    \setlength{\tabcolsep}{3pt}
    \renewcommand{\arraystretch}{0.95}
    \begin{threeparttable}
    \caption{Binary Classification Comparison of  the Proposed SA, JS, and SA-JS WGAN-GPs for Different IDS models}
    \label{tab.3}
    \begin{tabular}{llc ccc ccc}
        \toprule
        \multirow{2}{*}{IDS Model} & \multirow{2}{*}{WGAN Variant} & \multirow{2}{*}{Acc.} & \multicolumn{3}{c}{Normal} & \multicolumn{3}{c}{Abnormal} \\
        \cmidrule(lr){4-6} \cmidrule(lr){7-9}
         &  &  & Rec. & Prec. & F1-score & Rec. & Prec. & F1-score \\
        \midrule
        \multirow{4}{*}{\textbf{SVM}} 
         & WGAN-GP          &\textbf{75.1}$\pm$0.0 & 88.1$\pm$0.0 & 65.7$\pm$0.0 & 75.3$\pm$0.0 & 65.2$\pm$0.0 & 87.9$\pm$0.0 & 74.9$\pm$0.0 \\
         & SA-WGAN-GP       & 73.3$\pm$0.0  & 93.6$\pm$0.0 & 62.7$\pm$0.0 & 73.3$\pm$0.0 & 57.9$\pm$0.0 & 92.2$\pm$0.0 & 71.2$\pm$0.0 \\
         & JS-WGAN-GP       & 73.1$\pm$0.0 & 98.2$\pm$0.0 & 61.8$\pm$0.0 & 73.1$\pm$0.0 & 54.2$\pm$0.0 & 97.5$\pm$0.0 & 69.6$\pm$0.0 \\
         & SA-JS-WGAN-GP    & 73.3$\pm$0.0  & 98.2$\pm$0.0 & 62.0$\pm$0.0 & 73.3$\pm$0.0 & 54.5$\pm$0.0 & 97.5$\pm$0.0 & 69.9$\pm$0.0 \\
        \addlinespace[1.5ex]
        \multirow{4}{*}{\textbf{DT}} 
         & WGAN-GP       & 78.8$\pm$1.0   & 91.1$\pm$1.8  & 70.4$\pm$2.2  & 80.9$\pm$1.3  & 49.5$\pm$3.5  & 95.3$\pm$1.6  & 80.3$\pm$2.1 \\
         & SA-WGAN-GP    & 78.3$\pm$0.8   & 91.7$\pm$0.8  & 68.5$\pm$1.0  & 78.5$\pm$0.7  & 68.2$\pm$1.5  & 91.6$\pm$0.2  & 78.2$\pm$1.0 \\
         & JS-WGAN-GP     & 81.3$\pm$0.8   & 94.7$\pm$0.4  & 73.6$\pm$0.9  & 82.9$\pm$0.6  & 74.3$\pm$1.1  & 94.9$\pm$0.4  & 83.4$\pm$0.7 \\
         & SA-JS-WGAN-GP   & \textbf{83.1}$\pm$0.6  & 90.0$\pm$1.1  & 77.0$\pm$0.5  & 83.0$\pm$0.4  & 79.7$\pm$0.7  & 91.4$\pm$0.8  & 85.1$\pm$0.3 \\
        \addlinespace[1.5ex]
        \multirow{4}{*}{\textbf{DNN}} 
         & WGAN-GP          & 75.7$\pm$6.0  & 76.5$\pm$9.1  & 83.1$\pm$11.4  & 78.4$\pm$2.0  & 76.5$\pm$9.1  & 83.1$\pm$11.4  & 78.4$\pm$2.0 \\
         & SA-WGAN-GP       & 78.5$\pm$1.5 & 97.2$\pm$0.2  & 67.4$\pm$1.5  & 79.6$\pm$1.0  & 64.3$\pm$2.4  & 96.9$\pm$0.2  & 77.3$\pm$1.8 \\
         & JS-WGAN-GP       & \textbf{82.3}$\pm$1.6  & 92.1$\pm$4.9  & 73.8$\pm$3.4  & 81.8$\pm$1.1  & 74.9$\pm$5.5  & 93.0$\pm$3.4  & 82.7$\pm$2.2 \\
         & SA-JS-WGAN-GP    & 81.6$\pm$3.0 & 89.8$\pm$9.7  & 73.7$\pm$3.1  & 80.6$\pm$4.0  & 75.4$\pm$5.7  & 91.6$\pm$6.7  & 82.3$\pm$2.5 \\
        \addlinespace[1.5ex]
        \multirow{4}{*}{\textbf{CNN}} 
         & WGAN-GP          & 76.6$\pm$5.1 & 62.7$\pm$10.6  & 96.5$\pm$2.0  & 75.3$\pm$7.5  & 96.7$\pm$2.4  & 66.8$\pm$5.9  & 78.8$\pm$3.4 \\
         & SA-WGAN-GP       & 80.9$\pm$1.3  & 95.9$\pm$0.9  & 70.5$\pm$1.4  & 81.2$\pm$0.9  & 69.6$\pm$2.1  & 95.7$\pm$0.8  & 80.5$\pm$1.4 \\
         & JS-WGAN-GP       & 81.8$\pm$2.0   & 96.6$\pm$0.5  & 71.5$\pm$2.4  & 82.2$\pm$1.5  & 70.8$\pm$3.5  & 96.5$\pm$0.4  & 81.6$\pm$2.3 \\
         & SA-JS-WGAN-GP    & \textbf{81.9}$\pm$1.4 & 95.8$\pm$1.7  & 71.6$\pm$1.6  & 81.9$\pm$1.3  & 71.2$\pm$2.1  & 95.8$\pm$1.7  & 81.6$\pm$1.6 \\
        \addlinespace[1.5ex]
        \multirow{4}{*}{\textbf{LSTM}} 
         & WGAN-GP          & 81.4$\pm$0.7  & 97.4$\pm$0.2  & 72.7$\pm$2.4  & 82.1$\pm$0.7  & 72.4$\pm$3.4  & 96.9$\pm$0.5  & 82.8$\pm$2.1 \\
         & SA-WGAN-GP       & 80.9$\pm$1.2 & 96.5$\pm$1.2  & 70.0$\pm$1.3  & 81.5$\pm$0.8  & 69.6$\pm$2.1  & 95.7$\pm$0.8  & 80.5$\pm$1.4 \\
         & JS-WGAN-GP       & 80.6$\pm$1.7  & 95.4$\pm$1.8  & 70.4$\pm$2.2  & 80.9$\pm$1.3  & 69.5$\pm$3.5  & 95.3$\pm$1.6  & 80.3$\pm$2.1 \\
         & SA-JS-WGAN-GP    & \textbf{82.0}$\pm$3.3 & 96.2$\pm$1.8  & 71.9$\pm$3.7  & 82.2$\pm$2.6  & 71.3$\pm$5.1  & 96.1$\pm$1.9  & 81.8$\pm$3.6 \\
        \bottomrule
    \end{tabular}
    \begin{tablenotes}\footnotesize
        \item \textbf{Note:} The best average accuracy of different WGAN-GP variants under each IDS model is highlighted in bold.
    \end{tablenotes}
    \end{threeparttable}

\end{table}

\begin{figure}[!ht]
    \centering
    \begin{subfigure}[b]{0.9\textwidth}
        \centering
        \includegraphics[height=4.0cm]{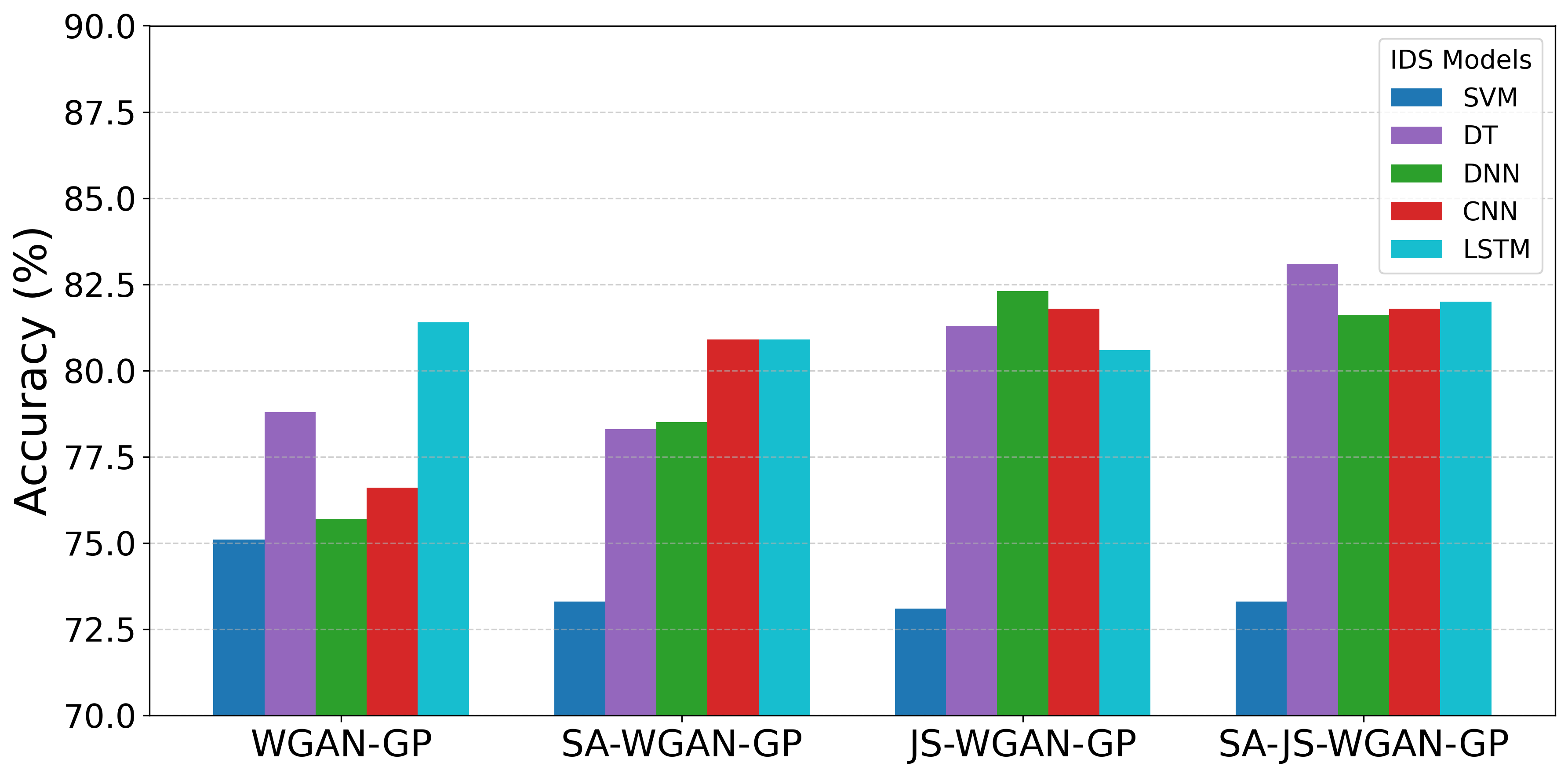}
        \caption{}
        \label{fig:ids_accuracy_line_chart}
    \end{subfigure}
    
    \vskip\baselineskip  

    \begin{subfigure}[b]{0.9\textwidth}
        \centering
        \includegraphics[height=4.0cm]{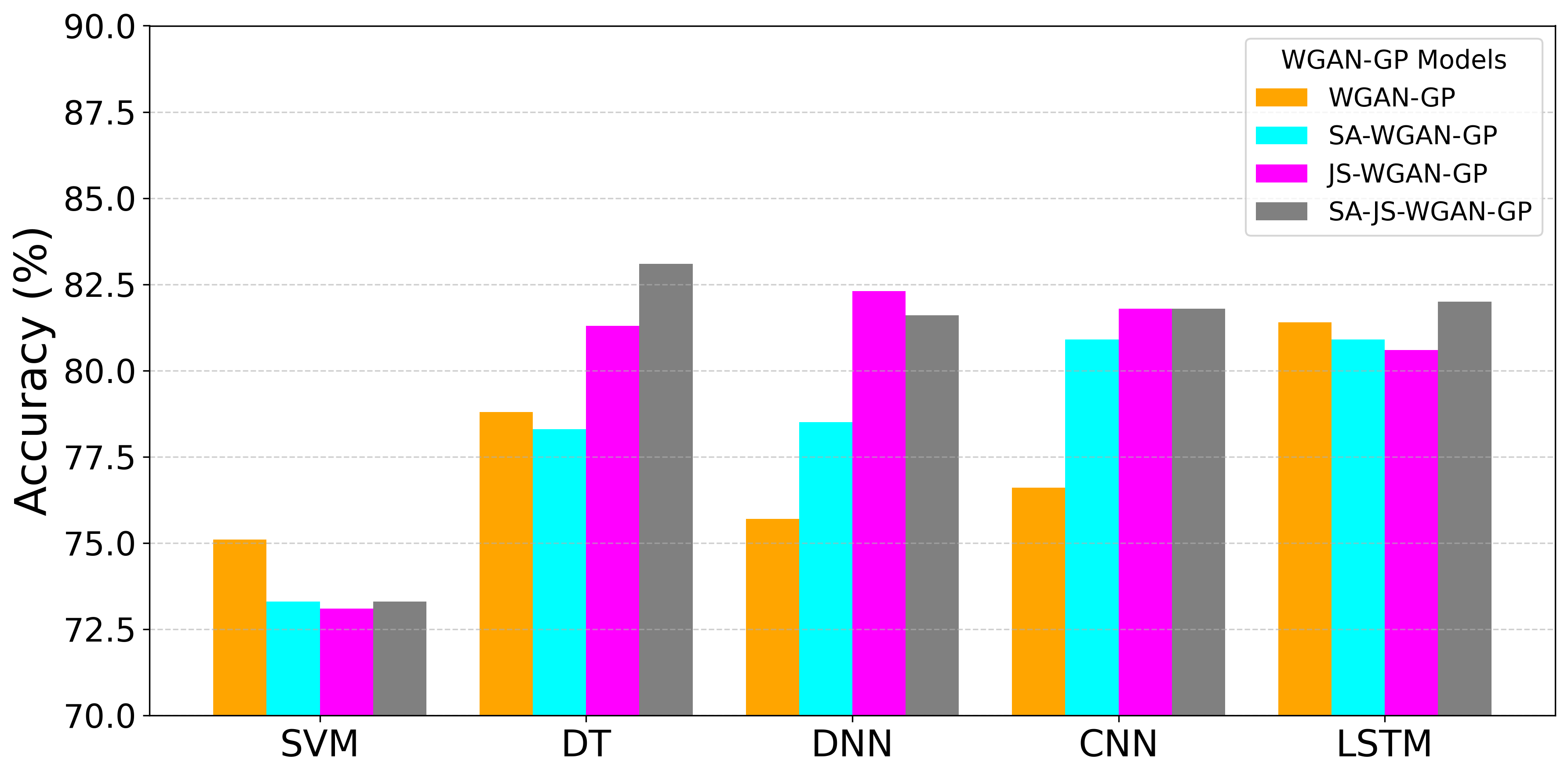}
        \caption{}
        \label{fig:wgan_gp_accuracies_line_chart}
    \end{subfigure}
    
    \caption{Binary classification average accuracy of various IDS models with different WGAN-GP variants. (a) Average accuracy of IDS models across different WGAN-GP variants. (b) Average accuracy of WGAN-GP variants across different IDS models.}
    \label{fig.4}
\end{figure}

The preprocessed NSL-KDD dataset was used for binary classification experiments, in which four types of attack traffic were treated as abnormal samples. Tables \ref{tab.2} and \ref{tab.3} list the binary classification outcomes of the five different IDS models, with each IDS model being trained on each of the four different combined datasets separately. Table \ref{tab.2} focuses on the comparison of different IDS models with a specific WGAN-GP model, whilst Table \ref{tab.3} is used in the comparison of different WGAN-GP models for a specific IDS model. In terms of detection accuracy, the tables show that, in most cases, the IDS models trained with the proposed WGAN-GP variants are better than those trained with the baseline WGAN-GP model. This indicates that the proposed WGAN-GP models are able to improve the quality of the generated data compared to the baseline WGAN-GP model, thereby enhancing the accuracy of the IDS models. Specifically, it is noticeable that the DT, DNN, CNN, and LSTM-based IDS models improved their average accuracy by 4.3\%, 5.9\%, 5.2\%, and 0.6\% with the proposed SA-JS-WGAN-GP, respectively, from Table \ref{tab.3}. However, for the SVM model, the average accuracy decreased by approximately 2\% across the three proposed WGAN-GP variants. Figure \ref{fig.4} presents the average accuracy results for different IDS models. 

\subsection{Multi-Classification}

\begin{table}[!ht]
    \centering
    \scriptsize
    \setlength{\tabcolsep}{3pt}
    \renewcommand{\arraystretch}{0.95}
    \begin{threeparttable}
    \caption{Multi-Classification Results of NSL-KDD Dataset with Different IDS Models for Different WGAN-GP Variants (Part 1: Acc., DoS and Probe)}
    \label{tab4.1}
   
    \begin{tabular}{%
        l   
        l   
        l   
        ccc 
        ccc 
    }
        \toprule
        \multirow{2}{*}{\textbf{WGAN Variant}} & 
        \multirow{2}{*}{\textbf{IDS Model}} & 
        \multirow{2}{*}{\textbf{Acc.}} & 
        \multicolumn{3}{c}{\textbf{DoS}} & 
        \multicolumn{3}{c}{\textbf{Probe}} \\
        \cmidrule(lr){4-6} \cmidrule(lr){7-9}
         &  &  & \textbf{Rec.} & \textbf{Prec.} & \textbf{F1} &
                  \textbf{Rec.} & \textbf{Prec.} & \textbf{F1} \\
        \midrule
        \multirow{5}{2cm}{\textbf{WGAN-GP}}
         & SVM    & 71.8$\pm$0.0 & 70.5$\pm$0.0 & 96.4$\pm$0.0 & 81.5$\pm$0.0 & 61.3$\pm$0.0 & 69.4$\pm$0.0 & 65.1$\pm$0.0 \\
         & DT     & 73.4$\pm$0.5 & 84.8$\pm$2.1 & 95.6$\pm$0.6 & 89.9$\pm$1.2 & 72.8$\pm$1.4 & 76.3$\pm$0.4 & 74.5$\pm$0.9 \\
         & DNN    & 77.7$\pm$1.9  & 82.6$\pm$2.5 & 90.5$\pm$4.5 & 86.3$\pm$1.3 & 74.2$\pm$5.7 & 71.0$\pm$2.6 & 72.5$\pm$3.2 \\
         & CNN    & \textbf{79.1}$\pm$1.0 & 84.0$\pm$1.5 & 91.8$\pm$2.4 & 87.7$\pm$1.5 & 82.9$\pm$1.4 & 72.7$\pm$3.7 & 77.4$\pm$2.1  \\
         & LSTM   & 78.0$\pm$5.2 & 80.7$\pm$2.5 & 90.9$\pm$3.9 & 85.5$\pm$2.7 & 79.2$\pm$7.9 & 72.0$\pm$4.2 & 75.3$\pm$5.5  \\
        \addlinespace[1.5ex]
        \multirow{5}{2cm}{\textbf{SA-WGAN-GP}}
         & SVM    & 73.1$\pm$0.0 & 71.1$\pm$0.0 & 94.6$\pm$0.0 & 81.2$\pm$0.0 & 71.2$\pm$0.0 & 78.9$\pm$0.0 & 74.9$\pm$0.0 \\
         & DT     & 76.1$\pm$1.7  & 72.2$\pm$1.1 & 93.3$\pm$4.4 & 81.3$\pm$1.9 & 68.2$\pm$0.7 & 80.4$\pm$2.9 & 73.8$\pm$1.3 \\
         & DNN    & 75.3$\pm$0.6 & 77.6$\pm$0.4 & 95.7$\pm$0.7 & 85.7$\pm$0.4 & 63.0$\pm$2.9 & 83.8$\pm$1.7 & 71.9$\pm$2.0 \\
         & CNN    & 78.5$\pm$2.1 & 85.2$\pm$1.7 & 96.6$\pm$0.5 & 90.5$\pm$0.8 & 81.8$\pm$1.9 & 80.7$\pm$1.2 & 81.2$\pm$1.3 \\
         & LSTM   & \textbf{79.3}$\pm$2.2  & 84.8$\pm$5.5 & 95.2$\pm$2.2 & 89.6$\pm$3.3 & 81.0$\pm$3.8 & 48.5$\pm$13.3 & 59.7$\pm$12.9 \\
        \addlinespace[1.5ex]
       \multirow{5}{2cm}{\textbf{JS-WGAN-GP}}
         & SVM    & 74.4$\pm$0.0 & 75.2$\pm$0.0 & 97.2$\pm$0.0 & 84.8$\pm$0.0 & 69.1$\pm$0.0 & 90.7$\pm$0.0 & 69.1$\pm$0.0 \\
         & DT     & 79.5$\pm$0.9  & 86.8$\pm$0.4 & 95.2$\pm$0.5 & 90.8$\pm$0.3 & 82.1$\pm$1.1 & 81.6$\pm$1.7 & 81.9$\pm$0.9 \\
         & DNN    & 80.1$\pm$2.3  & 84.4$\pm$3.6 & 94.2$\pm$1.7 & 89.0$\pm$2.5 & 74.2$\pm$4.7 & 75.4$\pm$3.1 & 74.7$\pm$2.9 \\
         & CNN    & 81.3$\pm$1.3  & 86.6$\pm$3.0 & 95.8$\pm$1.3 & 91.0$\pm$1.8 & 82.4$\pm$3.2 & 80.7$\pm$1.1 & 81.5$\pm$1.8 \\
         & LSTM   & \textbf{84.3}$\pm$1.1 & 87.4$\pm$1.8 & 93.7$\pm$1.5 & 90.2$\pm$0.9 & 82.1$\pm$3.5 & 79.4$\pm$1.7 & 80.6$\pm$1.6\\
        \addlinespace[1.5ex]
        \multirow{5}{2cm}{\textbf{SA-JS-WGAN-GP}}
         & SVM    & 74.6$\pm$0.0 & 75.2$\pm$0.0 & 97.2$\pm$0.0 & 84.8$\pm$0.0 & 69.1$\pm$0.0 & 90.7$\pm$0.0 & 69.1$\pm$0.0 \\
         & DT     & 81.9$\pm$1.3 & 93.8$\pm$0.6 & 93.7$\pm$0.5 & 93.7$\pm$0.3 & 77.3$\pm$1.0 & 78.8$\pm$5.4 & 78.0$\pm$2.8 \\
         & DNN    & 82.7$\pm$0.7  & 84.6$\pm$1.9 & 94.8$\pm$2.1 & 89.4$\pm$1.0 & 69.2$\pm$4.5 & 75.8$\pm$2.9 & 72.2$\pm$2.4  \\
         & CNN    & 80.0$\pm$2.2  & 82.8$\pm$3.5 & 96.1$\pm$1.4 & 88.9$\pm$1.6 & 80.8$\pm$4.2 & 81.5$\pm$2.5 & 81.0$\pm$2.0 \\
         & LSTM   & \textbf{85.9}$\pm$0.4 & 90.7$\pm$1.7 & 93.5$\pm$2.6 & 92.0$\pm$1.1 & 71.2$\pm$1.3 & 81.7$\pm$1.7 & 76.1$\pm$1.1 \\
        \midrule
    \end{tabular}
    \begin{tablenotes}\footnotesize
      \item \textbf{Note:} “F1” = F1-score. The best average accuracy values of different IDS models for different WGAN-GP variants are highlighted in bold.
    \end{tablenotes}
    \end{threeparttable}
\end{table}

\begin{table}[!htbp]
    \centering
    \scriptsize
    \setlength{\tabcolsep}{3pt}
    \renewcommand{\arraystretch}{0.95}
    \begin{threeparttable}
    \caption{Multi-Classification Results of NSL-KDD Dataset with Different IDS Models for Different WGAN-GP Variants (Part 2: R2L and U2R)}
    \label{tab4.2}
    \begin{tabular}{%
        l   
        l   
        r   
        ccc 
        ccc 
    }
        \toprule
        \multirow{2}{*}{\textbf{WGAN Variant}} & 
        \multirow{2}{*}{\textbf{IDS Model}} & 
        \multirow{2}{*}{\phantom{Acc.} } & 
        \multicolumn{3}{c}{\textbf{R2L}} & 
        \multicolumn{3}{c}{\textbf{U2R}} \\
        \cmidrule(lr){4-6} \cmidrule(lr){7-9}
         &  &  & \textbf{Rec.} & \textbf{Prec.} & \textbf{F1} &
                  \textbf{Rec.} & \textbf{Prec.} & \textbf{F1} \\
        \midrule
        \multirow{5}{2cm}{\textbf{WGAN-GP}}
           & SVM    &\phantom{Acc.} & -     & -     & -     & -     & -     & -    \\
           & DT     &\phantom{Acc.} & 14.6$\pm$2.3  & 96.8$\pm$1.1  & 25.3$\pm$3.5   & 11.9$\pm$0.0  & 30.0$\pm$4.5  & 17.0$\pm$0.9 \\
           & DNN    &\phantom{Acc.} & 39.1$\pm$15.7  & 71.6$\pm$12.2  & 49.1$\pm$5.5  & 26.6$\pm$5.2  & 26.7$\pm$13.2  & 24.7$\pm$6.2 \\
           & CNN    &\phantom{Acc.} & 9.4$\pm$5.0  & 96.0$\pm$2.8  & 16.8$\pm$8.1  & 34.8$\pm$2.9  & 23.4$\pm$7.7  & 27.3$\pm$5.5 \\
           & LSTM   &\phantom{Acc.} & 50.9$\pm$17.2  & 64.3$\pm$5.4  & 54.0$\pm$18.1  & 34.2$\pm$4.7  & 6.7$\pm$3.7  & 10.9$\pm$5.3 \\
        \addlinespace[1.5ex]
        \multirow{5}{2cm}{\textbf{SA-WGAN-GP}}
           & SVM    &\phantom{Acc.} & 3.9$\pm$0.0   & 84.3$\pm$0.0  & 7.5$\pm$0.0   & 10.4$\pm$0.0  & 43.8$\pm$0.0  & 16.9$\pm$0.0 \\
           & DT     &\phantom{Acc.} & 57.7$\pm$2.7  & 64.7$\pm$6.4  & 60.9$\pm$3.8  & 17.2$\pm$11.9  & 10.7$\pm$3.8  & 11.6$\pm$4.8 \\
           & DNN    &\phantom{Acc.} & 6.6$\pm$2.7   & 86.3$\pm$8.6  & 12.2$\pm$4.8   & 19.7$\pm$3.3  & 61.8$\pm$10.4  & 29.7$\pm$4.5 \\
           & CNN    &\phantom{Acc.} & 14.2$\pm$7.7   & 95.1$\pm$2.5  & 24.0$\pm$11.3   & 21.1$\pm$7.1  & 55.1$\pm$17.7  & 30.9$\pm$7.6 \\
           & LSTM   &\phantom{Acc.} & 33.6$\pm$15.9  & 68.9$\pm$5.4  & 42.1$\pm$19.0  & 33.6$\pm$2.7  & 32.9$\pm$21.9  & 29.8$\pm$10.5 \\
        \addlinespace[1.5ex]
        \multirow{5}{2cm}{\textbf{JS-WGAN-GP}}
           & SVM   &\phantom{Acc.} & 0.2$\pm$0.0   & 40.0$\pm$0.0  & 0.4$\pm$0.0   & 6.0$\pm$0.0   & 8.2$\pm$0.0   & 6.9$\pm$0.0   \\
           & DT    &\phantom{Acc.} & 42.2$\pm$6.8  & 65.9$\pm$12.4  & 51.3$\pm$8.2  & 15.6$\pm$1.7  & 3.8$\pm$3.2   & 5.6$\pm$2.9   \\
           & DNN   &\phantom{Acc.} & 36.4$\pm$9.3 & 74.4$\pm$4.5  & 48.3$\pm$8.8  & 28.2$\pm$7.0  & 20.8$\pm$6.7  & 24.8$\pm$7.0 \\
           & CNN   &\phantom{Acc.} & 17.7$\pm$4.1  & 94.1$\pm$8.6  & 29.5$\pm$5.9  & 21.9$\pm$6.8  & 63.4$\pm$15.2  & 32.0$\pm$8.7  \\
           & LSTM  &\phantom{Acc.} & 73.5$\pm$6.7  & 70.4$\pm$2.0  & 71.7$\pm$2.9  & 39.0$\pm$7.3  & 8.6$\pm$2.9   & 13.9$\pm$4.0 \\
        \addlinespace[1.5ex]
        \multirow{5}{2cm}{\textbf{SA-JS-WGAN-GP}}
           & SVM   &\phantom{Acc.} & 0.2$\pm$0.0   & 33.3$\pm$0.0  & 0.5$\pm$0.0   & 34.3$\pm$0.0  & 63.9$\pm$0.0  & 44.7$\pm$0.0 \\
           & DT    &\phantom{Acc.} & 55.6$\pm$0.6  & 79.1$\pm$0.2  & 65.3$\pm$0.5  & 17.7$\pm$2.0  & 5.2$\pm$2.5   & 7.6$\pm$2.9  \\
           & DNN   &\phantom{Acc.} & 51.2$\pm$6.4 & 84.8$\pm$4.1  & 63.5$\pm$4.8  & 23.2$\pm$10.3  & 44.7$\pm$14.7  & 27.4$\pm$8.9 \\
           & CNN   &\phantom{Acc.} & 15.8$\pm$7.2  & 96.3$\pm$2.5  & 26.6$\pm$10.1  & 17.1$\pm$5.4  & 42.1$\pm$11.1  & 23.3$\pm$5.4 \\
           & LSTM  &\phantom{Acc.} & 70.4$\pm$6.7  & 77.0$\pm$1.9  & 73.4$\pm$4.0  & 25.9$\pm$9.1  & 16.7$\pm$8.6 & 18.6$\pm$5.1 \\
        \bottomrule
    \end{tabular}
    \begin{tablenotes}\footnotesize
      \item \textbf{Note:} “F1” = F1-score.
    \end{tablenotes}
    \end{threeparttable}
\end{table}

\begin{figure}
    \centering
    \begin{subfigure}[b]{0.9\textwidth}
        \centering
        \includegraphics[height=4.0cm]{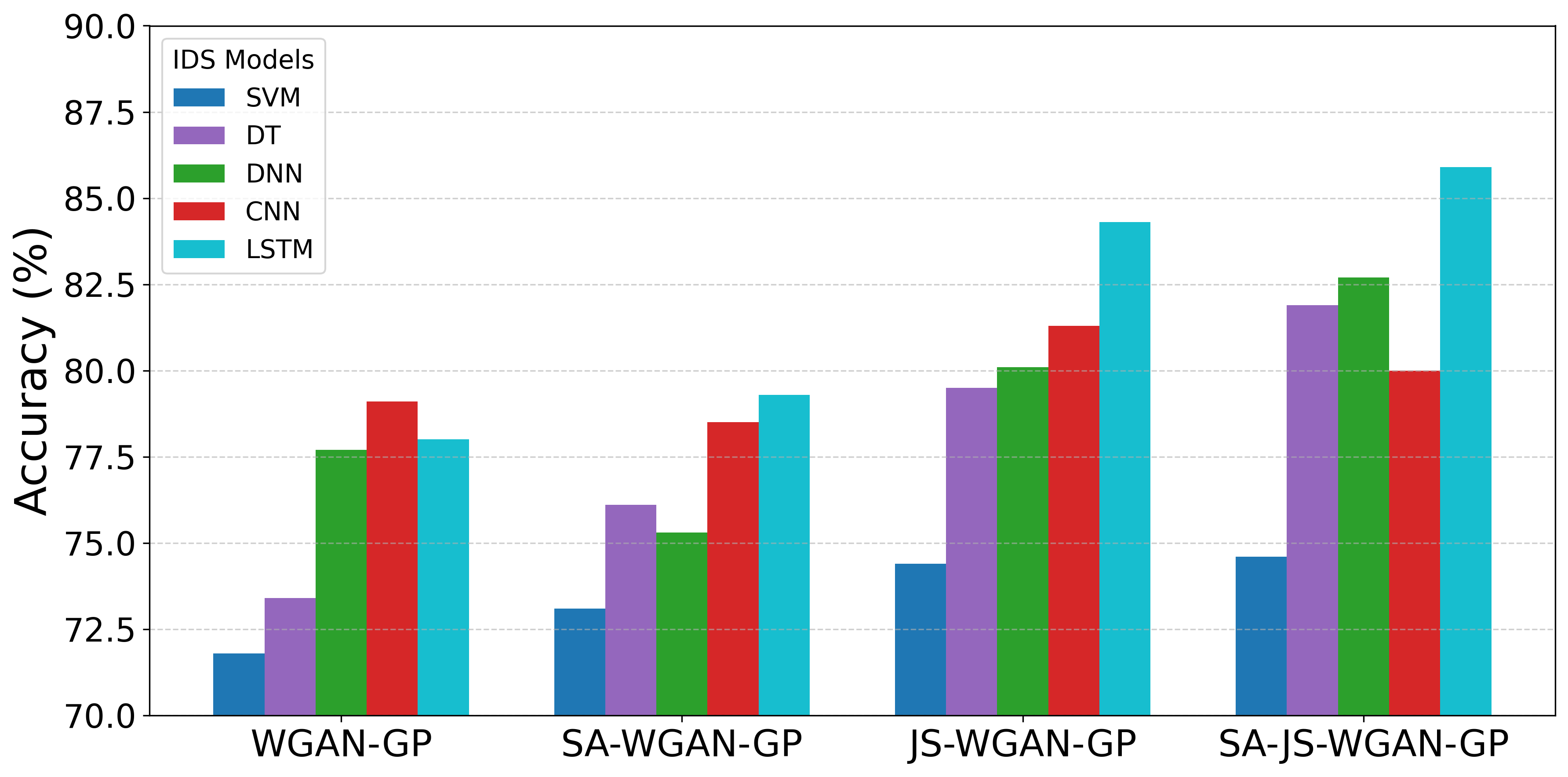}
        \caption{}
        \label{fig:ids_accuracy_line_chart}
    \end{subfigure}
    
    \vskip\baselineskip  

    \begin{subfigure}[b]{0.9\textwidth}
        \centering
        \includegraphics[height=4.0cm]{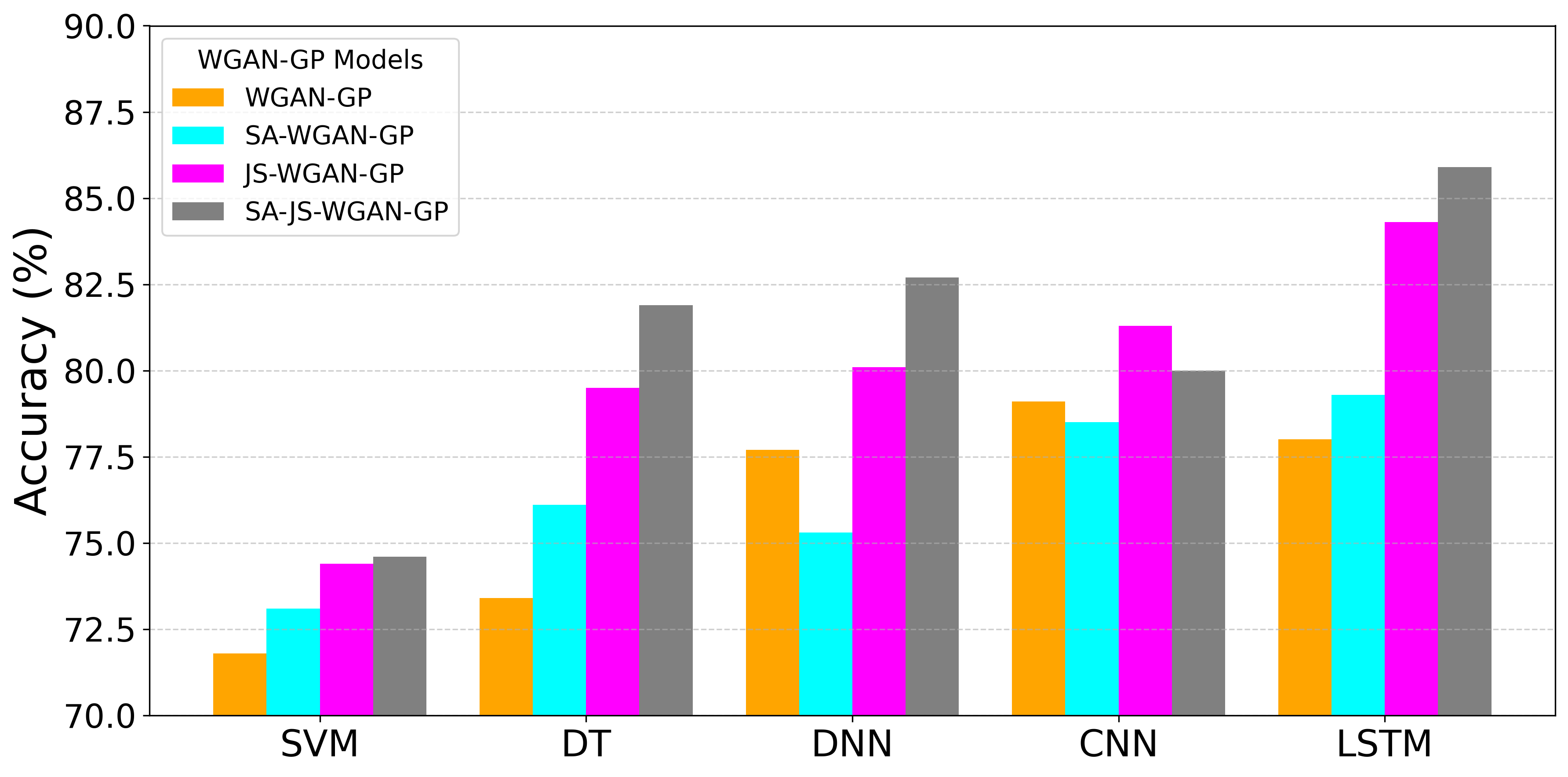}
        \caption{}
        \label{fig:wgan_gp_accuracies_line_chart}
    \end{subfigure}
    
    \caption {Multi-classification average accuracy of various IDS models with different WGAN-GP variants. (a) Average accuracy of IDS models across different WGAN-GP variants; (b) Average accuracy of WGAN-GP variants across different IDS models}
    \label{fig.5}
\end{figure}

Unlike the binary classification test, this experiment reports multi-classi\-fication results across five data categories, along with the corresponding recall, precision, and F1-score for four distinct attack types. Tables \ref{tab4.1} and \ref{tab4.2} present the performance of the five IDS models trained with the same combined dataset under the same WGAN-GP framework, whilst Tables \ref{tab5.1} and \ref{tab5.2} list the classification results of the same IDS model trained on different combined datasets generated by different WGAN-GP models, respectively. The results in Tables \ref{tab4.1}, \ref{tab4.2}, \ref{tab5.1}, and \ref{tab5.2} show that, in terms of accuracy, all three proposed WGAN-GP models outperform the baseline model in most scenarios. Comparing the highest average accuracy of the three proposed WGAN-GP variants with that of the baseline WGAN-GP for each IDS model in Table \ref{tab5.1}, the IDS models based on SVM, DT, DNN, CNN, and LSTM achieve maximum improvement of 2.8\%, 8.5\%, 5\%, 2.2\%, and 7.9\%, respectively. Figure \ref{fig.5} shows the accuracy variation of the IDS models across the basic and proposed WGAN-GP variants. It clearly shows that DT, DNN, and LSTM-based IDS models achieve their highest multi-classification average accuracy with the proposed SA-JS-WGAN-GP model.

\begin{table}[!htbp]
    \centering
    \scriptsize
    \setlength{\tabcolsep}{3pt}
    \renewcommand{\arraystretch}{0.95}
    \begin{threeparttable}
    \caption{Multi-Classification Results of NSL-KDD Dataset with Different WGAN-GP Variants for Different IDS Models (Part 1: Acc., DoS and Probe)}
    \label{tab5.1}
    \begin{tabular}{%
        l   
        l   
        l   
        ccc 
        ccc 
    }
        \toprule
        \multirow{2}{*}{\textbf{IDS Model}} & 
        \multirow{2}{*}{\textbf{WGAN Variant}} & 
        \multirow{2}{*}{\textbf{Acc.}} & 
        \multicolumn{3}{c}{\textbf{DoS}} & 
        \multicolumn{3}{c}{\textbf{Probe}} \\
        \cmidrule(lr){4-6} \cmidrule(lr){7-9}
         &  &  & \textbf{Rec.} & \textbf{Prec.} & \textbf{F1} &
                  \textbf{Rec.} & \textbf{Prec.} & \textbf{F1} \\
        \midrule
        \multirow{4}{*}{\textbf{SVM}}
         & WGAN-GP         & 71.8$\pm$0.0 & 70.5$\pm$0.0 & 96.4$\pm$0.0 & 81.5$\pm$0.0 & 61.3$\pm$0.0 & 69.4$\pm$0.0 & 65.1$\pm$0.0  \\
         & SA-WGAN-GP       & 73.1$\pm$0.0 & 71.1$\pm$0.0 & 94.6$\pm$0.0 & 81.2$\pm$0.0 & 71.2$\pm$0.0 & 78.9$\pm$0.0 & 74.9$\pm$0.0  \\
         & JS-WGAN-GP      & 74.4$\pm$0.0 & 75.2$\pm$0.0 & 97.2$\pm$0.0 & 84.8$\pm$0.0 & 69.1$\pm$0.0 & 90.7$\pm$0.0 & 69.1$\pm$0.0 \\
         & SA-JS-WGAN-GP    & {\textbf{74.6}}$\pm$0.0 & 75.2$\pm$0.0 & 97.2$\pm$0.0 & 84.8$\pm$0.0 & 69.1$\pm$0.0 & 90.7$\pm$0.0 & 69.1$\pm$0.0  \\
        \addlinespace[1.5ex]
        \multirow{4}{*}{\textbf{DT}} 
         & WGAN-GP          & 73.4$\pm$0.5 & 84.8$\pm$2.1 & 95.6$\pm$0.6 & 89.9$\pm$1.2 & 72.8$\pm$1.4 & 76.3$\pm$0.4 & 74.5$\pm$0.9  \\
         & SA-WGAN-GP       & 76.1$\pm$1.7  & 72.2$\pm$1.1 & 93.3$\pm$4.4 & 81.3$\pm$1.9 & 68.2$\pm$0.7 & 80.4$\pm$2.9 & 73.8$\pm$1.3  \\
         & JS-WGAN-GP       & 79.5$\pm$0.9  & 86.8$\pm$0.4 & 95.2$\pm$0.5 & 90.8$\pm$0.3 & 82.1$\pm$1.1 & 81.6$\pm$1.7 & 81.9$\pm$0.9 \\
         & SA-JS-WGAN-GP   & {\textbf{81.9}}$\pm$1.3 & 93.8$\pm$0.6 & 93.7$\pm$0.5 & 93.7$\pm$0.3 & 77.3$\pm$1.0 & 78.8$\pm$5.4 & 78.0$\pm$2.8   \\
        \addlinespace[1.5ex]
        \multirow{4}{*}{\textbf{DNN}} 
         & WGAN-GP          & 77.7$\pm$1.9  & 82.6$\pm$2.5 & 90.5$\pm$4.5 & 86.3$\pm$1.3 & 74.2$\pm$5.7 & 71.0$\pm$2.6 & 72.5$\pm$3.26  \\
         & SA-WGAN-GP       & 75.3$\pm$0.6 & 77.6$\pm$0.4 & 95.7$\pm$0.7 & 85.7$\pm$0.4 & 63.0$\pm$2.9 & 83.8$\pm$1.7 & 71.9$\pm$2.0 \\ 
         & JS-WGAN-GP       & 80.1$\pm$2.3  & 84.4$\pm$3.6 & 94.2$\pm$1.7 & 89.0$\pm$2.5 & 74.2$\pm$4.7 & 75.4$\pm$3.1 & 74.7$\pm$2.9  \\
         & SA-JS-WGAN-GP    & {\textbf{82.7}}$\pm$0.7  & 84.6$\pm$1.9 & 94.8$\pm$2.1 & 89.4$\pm$1.0 & 69.2$\pm$4.5 & 75.8$\pm$2.9 & 72.2$\pm$2.4  \\
         \addlinespace[1.5ex]
        \multirow{4}{*}{\textbf{CNN}} 
         & WGAN-GP          & 79.1$\pm$1.0 & 84.0$\pm$1.5 & 91.8$\pm$2.4 & 87.7$\pm$1.5 & 82.9$\pm$1.4 & 72.7$\pm$3.7 & 77.4$\pm$2.1   \\
         & SA-WGAN-GP        & 78.5$\pm$2.1 & 85.2$\pm$1.7 & 96.6$\pm$0.5 & 90.5$\pm$0.8 & 81.8$\pm$1.9 & 80.7$\pm$1.2 & 81.2$\pm$1.3  \\
         & JS-WGAN-GP       & {\textbf{81.3}}$\pm$1.3  & 86.6$\pm$3.0 & 95.8$\pm$1.3 & 91.0$\pm$1.8 & 82.4$\pm$3.2 & 80.7$\pm$1.1 & 81.5$\pm$1.8  \\
         & SA-JS-WGAN-GP    & 80.0$\pm$2.2  & 82.8$\pm$3.5 & 96.1$\pm$1.4 & 88.9$\pm$1.6 & 80.8$\pm$4.2 & 81.5$\pm$2.5 & 81.0$\pm$2.0  \\
        \addlinespace[1.5ex]
        \multirow{4}{*}{\textbf{LSTM}} 
         & WGAN-GP         & 78.0$\pm$5.2 & 80.7$\pm$2.5 & 90.9$\pm$3.9 & 85.5$\pm$2.7 & 79.2$\pm$7.9 & 72.0$\pm$4.2 & 75.3$\pm$5.5  \\
         & SA-WGAN-GP       & 79.3$\pm$2.2  & 84.8$\pm$5.5 & 95.2$\pm$2.2 & 89.6$\pm$3.3 & 81.0$\pm$3.8 & 48.5$\pm$13.3 & 59.7$\pm$12.9  \\
         & JS-WGAN-GP       & 84.3$\pm$1.1 & 87.4$\pm$1.8 & 93.7$\pm$1.5 & 90.2$\pm$0.9 & 82.1$\pm$3.5 & 79.4$\pm$1.7 & 80.6$\pm$1.6 \\
         & SA-JS-WGAN-GP     & {\textbf{85.9}}$\pm$0.4 & 90.7$\pm$1.7 & 93.5$\pm$2.6 & 92.0$\pm$1.1 & 71.2$\pm$1.3 & 81.7$\pm$1.7 & 76.1$\pm$1.1  \\
         \bottomrule
    \end{tabular}
    \begin{tablenotes}\footnotesize
      \item \textbf{Note:}  “F1” = F1-score. The best average accuracy values of different WGAN-GP variants under different IDS models are highlighted in bold.
    \end{tablenotes}
    \end{threeparttable}
\end{table}

\begin{table}[!htbp]
    \centering
    \scriptsize
    \setlength{\tabcolsep}{3pt}
    \renewcommand{\arraystretch}{0.95}
    \begin{threeparttable}
    \caption{Multi-Classification Results of NSL-KDD Dataset with Different WGAN-GP Variants for Different IDS Models (Part 2: R2L and U2R)}
    \label{tab5.2}
    \begin{tabular}{%
        l   
        l   
        l   
        ccc 
        ccc 
    }
        \toprule
        \multirow{2}{*}{\textbf{IDS Model}} & 
        \multirow{2}{*}{\textbf{WGAN Variant }} & 
        \multirow{1}{*}{\phantom{Acc.} } & 
        \multicolumn{3}{c}{\textbf{R2L}} & 
        \multicolumn{3}{c}{\textbf{U2R}} \\
        \cmidrule(lr){4-6} \cmidrule(lr){7-9}
         &  &  & \textbf{Rec.} & \textbf{Prec.} & \textbf{F1} &
                  \textbf{Rec.} & \textbf{Prec.} & \textbf{F1} \\
        \midrule
        \multirow{4}{*}{\textbf{SVM}}
         & WGAN-GP &\phantom{Acc.} & -     & -     & -     & -     & -     & -  \\
         & SA-WGAN-GP &\phantom{Acc.}& 3.9$\pm$0.0   & 84.3$\pm$0.0  & 7.5$\pm$0.0   & 10.4$\pm$0.0  & 43.8$\pm$0.0  & 16.9$\pm$0.0 \\
         & JS-WGAN-GP  &\phantom{Acc.} & 0.2$\pm$0.0   & 40.0$\pm$0.0  & 0.4$\pm$0.0   & 6.0$\pm$0.0   & 8.2$\pm$0.0   & 6.9$\pm$0.0  \\
         & SA-JS-WGAN-GP &\phantom{Acc.}  & 0.2$\pm$0.0   & 33.3$\pm$0.0  & 0.5$\pm$0.0   & 34.3$\pm$0.0  & 63.9$\pm$0.0  & 44.7$\pm$0.0 \\
        \addlinespace[1.5ex]
        \multirow{4}{*}{\textbf{DT}} 
         & WGAN-GP &\phantom{Acc.} & 14.6$\pm$2.3  & 96.8$\pm$1.1  & 25.3$\pm$3.5   & 11.9$\pm$0.0  & 30.0$\pm$4.5  & 17.0$\pm$0.9 \\
         & SA-WGAN-GP &\phantom{Acc.} & 57.7$\pm$2.7  & 64.7$\pm$6.4  & 60.9$\pm$3.8  & 17.2$\pm$11.9  & 10.7$\pm$3.8  & 11.6$\pm$4.8 \\
         & JS-WGAN-GP &\phantom{Acc.} & 42.2$\pm$6.8  & 65.9$\pm$12.4  & 51.3$\pm$8.2  & 15.6$\pm$1.7  & 3.8$\pm$3.2   & 5.6$\pm$2.9 \\
         & SA-JS-WGAN-GP &\phantom{Acc.} & 55.6$\pm$0.6  & 79.1$\pm$0.2  & 65.3$\pm$0.5  & 17.7$\pm$2.0  & 5.2$\pm$2.5   & 7.6$\pm$2.9 \\
        \addlinespace[1.5ex]
        \multirow{4}{*}{\textbf{DNN}} 
         & WGAN-GP &\phantom{Acc.} & 39.1$\pm$15.7  & 71.6$\pm$12.2  & 49.1$\pm$5.5  & 26.6$\pm$5.2  & 26.7$\pm$13.2  & 24.7$\pm$6.2  \\
         & SA-WGAN-GP &\phantom{Acc.} & 6.6$\pm$2.7   & 86.3$\pm$8.6  & 12.2$\pm$4.8   & 19.7$\pm$3.3  & 61.8$\pm$10.4  & 29.7$\pm$4.5 \\
         & JS-WGAN-GP &\phantom{Acc.} & 36.4$\pm$9.3 & 74.4$\pm$4.5  & 48.3$\pm$8.8  & 28.2$\pm$7.0  & 20.8$\pm$6.7  & 24.8$\pm$7.0  \\ 
         & SA-JS-WGAN-GP &\phantom{Acc.} & 51.2$\pm$6.4 & 84.8$\pm$4.1  & 63.5$\pm$4.8  & 23.2$\pm$10.3  & 44.7$\pm$14.7  & 27.4$\pm$8.9 \\
         \addlinespace[1.5ex]
        \multirow{4}{*}{\textbf{CNN}} 
        & WGAN-GP &\phantom{Acc.} & 9.4$\pm$5.0  & 96.0$\pm$2.8  & 16.8$\pm$8.1  & 34.8$\pm$2.9  & 23.4$\pm$7.7  & 27.3$\pm$5.5 \\
        & SA-WGAN-GP &\phantom{Acc.} & 14.2$\pm$7.7   & 95.1$\pm$2.5  & 24.0$\pm$11.3   & 21.1$\pm$7.1  & 55.1$\pm$17.7  & 30.9$\pm$7.6 \\
        & JS-WGAN-GP &\phantom{Acc.} & 17.7$\pm$4.1  & 94.1$\pm$8.6  & 29.5$\pm$5.9  & 21.9$\pm$6.8  & 63.4$\pm$15.2  & 32.0$\pm$8.7 \\
        & SA-JS-WGAN-GP &\phantom{Acc.} & 15.8$\pm$7.2  & 96.3$\pm$2.5  & 26.6$\pm$10.1  & 17.1$\pm$5.4  & 42.1$\pm$11.1  & 23.3$\pm$5.4 \\
        \addlinespace[1.5ex]
        \multirow{4}{*}{\textbf{LSTM}} 
         & WGAN-GP &\phantom{Acc.} & 50.9$\pm$17.2  & 64.3$\pm$5.4  & 54.0$\pm$18.1  & 34.2$\pm$4.7  & 6.7$\pm$3.7  & 10.9$\pm$5.3 \\
         & SA-WGAN-GP &\phantom{Acc.}  & 33.6$\pm$15.9  & 68.9$\pm$5.4  & 42.1$\pm$19.0  & 33.6$\pm$2.7  & 32.9$\pm$21.9  & 29.8$\pm$10.5 \\
         & JS-WGAN-GP &\phantom{Acc.}  & 73.5$\pm$6.7  & 70.4$\pm$2.0  & 71.7$\pm$2.9  & 39.0$\pm$7.3  & 8.6$\pm$2.9   & 13.9$\pm$4.0 \\
         & SA-JS-WGAN-GP &\phantom{Acc.} & 70.4$\pm$6.7  & 77.0$\pm$1.9  & 73.4$\pm$4.0  & 25.9$\pm$9.1  & 16.7$\pm$8.6 & 18.6$\pm$5.1 \\
         \bottomrule
    \end{tabular}
    \begin{tablenotes}\footnotesize
      \item \textbf{Note:}  “F1” = F1-score.
    \end{tablenotes}
    \end{threeparttable}
\end{table}

Overall, sample generation enhances the generalization ability of IDS models against different attacks. In multi-classification tasks, SA-JS-WGAN-GP achieves higher performance than both JS-WGAN-GP and SA-WGAN-GP. Although all three models improve IDS detection accuracy, SA-JS-WGAN-GP is more adequate for deployment in resource-rich environments, whereas JS-WGAN-GP is preferable in resource-constrained settings when considering both detection performance and computational cost.

\subsection{Comparison with Related Works}

 The performance of the proposed three WGAN-GP variants is compared with other state-of-the-art GAN-based methods for intrusion detection in Table \ref{tab.6}. To ensure a fair comparison, all experiments were conducted under identical settings for data preprocessing, training procedure, and evaluation.

\begin{table}[!ht]
    \centering
    \scriptsize
    \begin{threeparttable}
    \caption{Performance Comparison of the Proposed SA-, JS-, and SA-JS-WGAN-GP with the State-of-the-art GAN models on NSL-KDD Dataset Based on IDS Detection Accuracy for Binary Classification (B) and Multi-classification (M).}
    \label{tab.6}
    \begin{tabular}{l c c c}
        \toprule
        \textbf{GAN Models} & \textbf{Acc. (B) (\%)} & \textbf{Acc. (M) (\%)} & \textbf{Training Time (hours)}\\
        \midrule
        SYN-GAN\cite{RAHMAN2024101212} & 79.1$\pm$5.6 & \underline{85.1}$\pm$1.2 & 5.4 \\
        IGAN-IDS\cite{HUANG2020102177} & 81.5$\pm$4.2 & 79.8$\pm$1.4 & 2.3 \\
        AI-based NIDS\cite{9908159} & 79.2$\pm$7.8 & 79.8$\pm$0.5 & 2.8 \\
        VAE-GAN\cite{9516484} & {\textbf{83.3}}$\pm$0.9 & 81.5$\pm$3.1 & 3.9 \\
        Mts-dvGAN\cite{SUN2024103570} & 80.1$\pm$1.6 & 80.7$\pm$0.7 & 5 \\
        GANS-ECNN\cite{9850286} & 80.8$\pm$3.6 & 76.1$\pm$2.6 & 15 \\
        Multi-Critics WGAN-GP\cite{10577721} & 79.4$\pm$1.1 & 81.9$\pm$1.0 & 100.7 \\
        SAWGAN\cite{GU2023366} & 80.0$\pm$2.1 & 78.3$\pm$1.2 & 91.4 \\
        WGAN-GP & 81.4$\pm$0.7 & 79.1$\pm$5.2 & 6.3 \\
        SA-WGAN-GP & 80.9$\pm$1.3 & 79.3$\pm$2.2 & 25.8 \\
        JS-WGAN-GP & 82.3$\pm$1.6 & 84.6$\pm$1.1 & 14.7 \\
        SA-JS-WGAN-GP & \underline{83.1}$\pm$0.6 & {\textbf{85.9}}$\pm$0.4 & 64.4\\
        \bottomrule
       \end{tabular}
        \begin{tablenotes}\footnotesize
            \item \textbf{Note:}  The best values are highlighted in bold, and the second-best value is underlined.
        \end{tablenotes}
   \end{threeparttable}
\end{table}

As shown in Table \ref{tab.6}, the VAE-GAN achieves the best average binary accuracy, marginally exceeding SA-JS-WGAN-GP by 0.2\%. In contrast, SA-JS-WGAN-GP attains the best multi-classification average accuracy (86.3\%) among all compared methods, outperforming VAE-GAN (81.5\%) and all other GAN variants tested against. This shows that while VAE-GAN slightly edges out our approach on the binary task, SA-JS-WGAN-GP provides the strongest overall gains, highlighting its superiority and robustness in generating a variety of new network attack data to improve IDS models in detecting network intrusions. This further indicates that our model enhances the IDS models' generalization ability and can capture patterns of various attacks more effectively.

\subsection{Results of LOAO in Emulating Zero-day Attack Detection}

\begin{table}[!ht]
    \centering
    \scriptsize
    \setlength{\tabcolsep}{3pt}
    \renewcommand{\arraystretch}{0.95}
    \begin{threeparttable}
        \caption{Comparison of LOAO Results Across Different GAN and IDS Models in Emulation of Zero-day Attack Detection}
        \label{tab:LOAO}

        \setlength{\tabcolsep}{2pt}
        \renewcommand{\arraystretch}{0.95}
        \begin{tabularx}{\columnwidth}{%
            >{\raggedright\arraybackslash}p{3.8cm}  
            >{\raggedright\arraybackslash}p{1.5cm}  
            *{4}{>{\centering\arraybackslash}X}
        }
        \toprule
        \textbf{GAN model} & \textbf{IDS model} & \textbf{Acc} & \textbf{Macro-F1} & \textbf{AUROC} & \textbf{TPR@5\%FPR} \\
        \midrule

        \multirow{5}{*}{\textbf{Baseline (No GAN)}}
            & LSTM & 84.8$\pm$1.1 & 71.7$\pm$1.3 & 64.0$\pm$7.2 & 7.0$\pm$1.4 \\
            & CNN  & \textbf{85.1$\pm$0.3}& \textbf{75.1$\pm$1.1} & \textbf{72.7$\pm$1.2} & \textbf{29.5$\pm$8.7} \\
            & DNN  & 78.4$\pm$6.9 & 61.4$\pm$5.4 & 45.4$\pm$8.8 & 14.4$\pm$7.7\\
            & DT   & 75.4$\pm$17.6 & 56.4$\pm$11.0 & 50.1$\pm$0.1 & 0.2$\pm$0.2 \\
            & SVM  & 80.8$\pm$0.0 & 62.2$\pm$0.0 & 47.0$\pm$0.0 & 0.0$\pm$0.0 \\
        \midrule

        \multirow{5}{*}{\textbf{WGAN-GP}}
            & LSTM & 83.4$\pm$1.4 & 60.8$\pm$1.2 & 46.8$\pm$11.0 & 2.9$\pm$1.6 \\
            & CNN  & \textbf{85.4}$\pm$0.7 & \textbf{73.3$\pm$1.6} & \textbf{78.3$\pm$1.2} & 24.3$\pm$6.0 \\
            & DNN  & 83.4$\pm$1.6 & 65.7$\pm$2.9 & 47.0$\pm$11.0 & \textbf{30.0$\pm$20.0}\\
            & DT   & 77.7$\pm$16.5 & 59.6$\pm$8.3 & 50.0$\pm$0.0 & 0.0$\pm$0.0 \\
            & SVM  & 82.3$\pm$0.0 & 61.0$\pm$0.0 & 47.0$\pm$0.0 & 0.0$\pm$0.0 \\
        \midrule

        \multirow{5}{*}{\textbf{SA-WGAN-GP}}
            & LSTM & 84.0$\pm$1.4 & 60.5$\pm$1.0 & 51.7$\pm$9.5 & 6.8$\pm$5.8 \\
            & CNN  & \textbf{85.4}$\pm$0.4 & \textbf{73.6$\pm$1.5} & \textbf{75.2$\pm$2.4} & 31.8$\pm$4.6 \\
            & DNN  & 77.9$\pm$8.6 & 61.6$\pm$7.2 & 49.6$\pm$11.0 & \textbf{38.1$\pm$23.0} \\
            & DT   & 58.8$\pm$13.6 & 51.8$\pm$6.4 & 50.0$\pm$0.0 & 0.0$\pm$0.0 \\
            & SVM  & 81.1$\pm$0.0 & 69.1$\pm$0.0 & 71.0$\pm$0.0 & 0.0$\pm$0.0 \\
        \midrule

        \multirow{5}{*}{\textbf{JS-WGAN-GP}}
            & LSTM & 84.0$\pm$1.5 & 61.5$\pm$1.8 & 47.4$\pm$10.1 & 6.2$\pm$2.4 \\
            & CNN  & \textbf{86.4$\pm$0.6} & \textbf{74.2$\pm$1.8} & \textbf{74.9$\pm$1.8} & \textbf{42.1$\pm$10.4} \\
            & DNN  & 84.0$\pm$1.5 & 67.3$\pm$4.1 & 55.6$\pm$6.9 & 29.4$\pm$16.5 \\
            & DT   & 56.2$\pm$12.9 & 50.1$\pm$6.2 & 50.5$\pm$0.5 & 1.0$\pm$1.0 \\
            & SVM  & 85.0$\pm$0.0 & \textbf{74.2$\pm$0.0} & 59.0$\pm$0.0 & 0.0$\pm$0.0 \\
        \midrule

        \multirow{5}{*}{\textbf{SA-JS-WGAN-GP}}
            & LSTM & 84.1$\pm$1.2 & 72.3$\pm$1.4 & 66.4$\pm$5.4 & 19.3$\pm$4.0 \\
            & CNN  & \textbf{86.3$\pm$1.1} & \textbf{73.9$\pm$2.0} & \textbf{80.0$\pm$3.3} & \textbf{36.4$\pm$7.3} \\
            & DNN  & 82.6$\pm$1.8 & 65.4$\pm$2.7 & 60.4$\pm$6.8 & 13.0$\pm$7.6 \\
            & DT   & 78.1$\pm$1.7 & 60.0$\pm$3.5 & 50.1$\pm$0.1 & 0.1$\pm$0.2 \\
            & SVM  & 85.3$\pm$0.0 & 67.3$\pm$0.0 & 69.8$\pm$0.0 & 28.5$\pm$0.0 \\
        \bottomrule
        \end{tabularx}

        \begin{tablenotes}\footnotesize
          \item \textbf{Note:} “ACC” = Known-only Acc. For each GAN model, including the Baseline, the best value of each evaluation metric across the five IDS classifiers is highlighted in bold.
        \end{tablenotes}
    \end{threeparttable}
\end{table}

We emulated zero-day attack detection on NSL-KDD using an LOAO experiment, where the entire R2L class is removed from all GAN and IDS models (including the Baseline model without GAN) training and validation stages. To address the class imbalance issues among the seen classes, only the minority classes are augmented by adding 20K synthetic Probe samples and 10K synthetic U2R samples, while the Normal and DoS samples remain unaltered, and the samples of R2L are excluded from both training and synthesis. IDS evaluation is conducted on the KDDTest, which includes the R2L attack class. In this way, the R2L samples are used as unseen attacks on the IDS models. The Known-only Accuracy and Known Macro-F1 over the seen classes are reported, along with R2L AUROC and R2L TPR@5\%FPR. A higher value of AUROC indicates a better discrimination ability. 

Table \ref{tab:LOAO} lists the LOAO results in emulation of zero-day attack detection. In terms of Known-only Accuracy, all WGAN-GP variants outperform the Baseline performance, with the CNN-based IDS model under JS-WGAN-GP achieving the highest value of 86.4\%. For Known Macro-F1, all  WGAN-GPs failed to surpass the Baseline performance, which was due to the distributional mismatch between the generated synthetic data and the real training distribution, thereby limiting the improvement in classifier generalization. Regarding AUROC, all four WGAN-GP variants exceed the Baseline performance, and the CNN-based IDS model in SA-JS-WGAN-GP  increases by 7.3\%, reaching 80.0\%. For TPR@5\%FPR, the CNN-based IDS model with JS-WGAN-GP surpasses the Baseline performance by 12.6\%, achieving 42.1\%. Overall, for the five IDS models evaluated for zero-day attack detection, SA-JS-WGAN-GP achieves the best balance, with an average AUROC of 66.1\% and TPR@5\%FPR of 19.5\%, improving by 10.3\% and 9.3\% over the Baseline. Compared with the Baseline, all four WGAN-GP models enhance unseen (zero-day) attack detectability in at least one metric, although improvements vary across metrics and IDS models, with SA-JS-WGAN-GP demonstrating the strongest robustness.

The augmentation strategy densifies and smooths the distribution of known classes, reducing overconfidence of IDS models in unknown samples. As a result, the unseen attack class R2L achieves a higher average AUROC in comparison with the Baseline, meaning the IDS models can discriminate unseen attacks better with the proposed SA-, JS-, and SA-JS-WGAN-GPs. In the LOAO evaluation, the CNN-based IDS model demonstrates the best performance, and SVM and DT are more sensitive to distribution shifts. Overall, under strict R2L removal with minority-class augmentation, SA-JS-WGAN-GP improves zero-day attack detectability without sacrificing performance on known classes.

\subsection{Further Discussion}

The SA mechanism allows an ML model to assign different importance factors to distinct parts of the input, enhancing the representation of key features. This selective attention enables the proposed WGAN-GP models to better distinguish between attacks and normal data by paying more attention to critical features and their relationships in the data. In addition, the JS divergence can effectively quantify the difference between the distribution of real and generated data, reducing model collapse, which is a common problem in generative models. Hence, using JS divergence during training, the proposed WGAN-GP models help generate more diverse synthetic data. By maintaining a higher diversity in synthetic data, the proposed WGAN-GP variants can better approximate the distribution of real-world data, thereby improving their generalization capabilities.

CNN excels at learning spatial features, which is advantageous when analyzing network traffic data, as relationships between data points may have a spatial component. Although the NSL-KDD dataset does not contain explicit temporal features, sequences of network events can be modeled by LSTM, capturing dependencies between related events to improve the classification accuracy of the LSTM-based IDS model. DNN provides hierarchical learning, enabling the IDS model to capture high-level and low-level patterns, critical for distinguishing complex attacks from normal traffic. DT also allows the model to identify feature interactions hierarchically, making it particularly suitable for interpreting complex datasets and providing more nuanced distinctions between different types of attacks.  While SVM is effective for classification, it struggles with large and high-dimensional datasets that are common in intrusion detection \cite{ZOUHRI2022115691}. One reason for this is that augmentation increases intra-class support and changes the decision boundary, moving beyond a single max-margin hyperplane \cite{jimaging7090177}, especially when feature scaling or kernel width is imperfect. In contrast, deep models adapt their representations, while traditional SVMs work with fixed features \cite{dablain2022efficientaugmentationimbalanceddeep}. This makes SVM less flexible when handling complex data and highlights the need for more adaptable models in intrusion detection. In binary classification experiments, all attack data were labeled as anomalies. This approach enriches the training set, but introduces more variability, which may blur the decision boundary between normal and attack classes. The wide variety of attacks in the synthetic data leads to similar features between normal and attack data, which can negatively affect the model's classification ability.

Data normalization and the `Tanh' activation function are used in all WGAN-GP variants to normalize the feature range to [-1.0, 1.0]. This ensures that all input features contribute effectively to the learning process. This prevents certain features from dominating due to their large data values, resulting in poor model performance. The smoother gradients provided by the `Tanh' activation function facilitate faster convergence and more stable learning, ensuring that the model can effectively learn the underlying data distribution and generate synthetic data closer to real data.

\begin{table}[!ht]
  \centering
  \scriptsize
  \begin{threeparttable}
    \caption{kNN/MMD Alignment by Different WGAN-GP Variants (Five Classes)}
    \label{tab:macro_knn_mmd}
    \setlength{\tabcolsep}{2.2pt}
    \renewcommand{\arraystretch}{0.95}
    \begin{tabularx}{\columnwidth}{%
      >{\raggedright\arraybackslash}p{3cm}
      *{5}{>{\centering\arraybackslash}X}
    }
      \toprule
      Model & frac\_lt\_eps & Macro kNN p50 & Macro kNN p95 & Macro MMD$^2$ (clip 0) & ALL MMD$^2$ \\
      \midrule
      WGAN-GP       & 0.0 & 0.148 & 0.686 & 0.000512 & 0.023734 \\
      SA-WGAN-GP    & 0.0 & 0.309 & 1.312 & 0.002859 & 0.021832 \\
      JS-WGAN-GP    & 0.0 & 7.824 & 8.548 & 0.379318 & 0.285558 \\
      SA-JS-WGAN-GP & 0.0 & 8.216 & 8.885 & 0.622680 & 0.343730 \\
      \bottomrule
    \end{tabularx}
    \begin{tablenotes}\footnotesize
      \item \textbf{Note:} Macro averages are unweighted means over \{Normal, DoS, Probe, U2R, R2L\}. ALL is computed on pooled real vs pooled synthetic samples (not an average). frac\_lt\_eps represents similarity. Macro MMD$^2$ is the unweighted average of each class's MMD$^2$. ALL MMD$^2$ are computed based on the combined real samples and the combined synthetic samples.
    \end{tablenotes}
  \end{threeparttable}
\end{table}

The alignment between synthetic and real training data across five classes is evaluated using three metrics as follows. 
\begin{itemize}
\item kNN $p50/p95$: the median and 95th-percentile nearest-neighbor distance from synthetic to real, lower values indicate the synthetic data is closer to the real samples, i.e., less distributed and small coverage; $p95$ reflects boundary coverage. 
\item $\mathrm{MMD}^2$ with an RBF kernel, smaller values indicate less distributional discrepancy, and small negative unbiased estimates are set to zero.
\item frac\_lt\_eps represents the proportion of synthetic points that lie within a small adaptive threshold, serving as a proxy for near-duplicate samples.
\end{itemize}

Across Table~\ref{tab:macro_knn_mmd}, WGAN-GP and SA-WGAN-GP show the tightest alignment (overall $\mathrm{MMD}^2 \approx 0.02$ and smaller kNN quantiles), indicating samples concentrated in high-density regions. JS-WGAN-GP and SA-JS-WGAN-GP yield larger $\mathrm{MMD}^2$ ($\approx 0.29$--$0.34$) and higher kNN quantiles ($p50 \approx 7$--$8$, $p95 \approx 8$--$9$), showing broader coverage of low-density and boundary areas. All models have frac\_lt\_eps $=0$, indicating no evidence of memorization. In summary, WGAN-based Gs favour high fidelity matching to the training distribution, whereas JS-WGAN-GP and SA-JS-WGAN-GP provide wider coverage that can support robustness by exposing classifiers to boundary and rare patterns.

Combining synthetic and original data can benefit the IDS models by enhancing their generalization ability. Synthetic data mimicking real data distribution provides additional training samples covering a wider range of potential attack scenarios. In particular, the superior performance of SA-JS-WGAN-GP carries important implications for zero-day attack detection in Internet of Things (IoT) and 5G networks. The proposed generative models can be used to expand training data in these scenarios, helping IDS adapts to complex network traffic, such as in the encrypted domain. At the same time, the SA mechanism can capture long-range dependencies and is suitable for high-dimensional multimodal data. Enhanced detection of rare and previously unseen attacks can lower the incidence of false alarms and strengthen defenses against advanced intrusions. This diverse dataset enables the IDS models to learn a more comprehensive decision boundary, which is critical for detecting zero-day attacks.

\subsection{Study Limitations}

WGAN-GP models with the SA mechanism and JS divergence have shown their potential in improving the detection of zero-day attacks for ML-based IDS systems. There remain some limitations in the work.

The proposed WGAN-GP variants are evaluated with the NSL-KDD dataset. Although the NSL-KDD dataset is widely used in the literature, it only partially represents real-world network traffic diversity. It is important to note that the performance of our models may vary when applied to more complex datasets. The integration of the SA mechanism and JS divergence leads to an increase in computational costs, which poses a significant challenge to the implementation of the proposed WGAN-GP variants in resource-constrained systems. Feature scaling and normalization for data preprocessing underscore the need for more advanced data filtering and processing methods to improve model performance further.

The augmented data generated by the SA-JS-WGAN-GP model enhances the generalization and robustness of most IDS models, including SVM, DT, DNN, and LSTM-based models, by providing a richer and more diverse set of samples. These IDS models leverage enhanced data diversity to boost performance in detecting attacks. However, CNNs, which excel at extracting local spatial features \cite{RAJESHKANNA2021107132}, appear less compatible with the SA mechanism and the additional complexity introduced by JS divergence, resulting in a slight decrease in average accuracy (1.3\%) for multi-classification tasks compared to JS-WGAN-GP. This suggests that while the SA-JS-WGAN-GP model effectively captures global dependencies, its added complexity is inconsistent with the core focus of some IDS models, such as CNNs.

\section{Conclusion}

As cyber threats become more sophisticated and frequent, preventing zero-day attacks has become a key challenge. This research evaluates the effectiveness of three GAN models, namely the SA-WGAN-GP, JS-WGAN-GP, and SA-JS-WGAN-GP, derived from the basic WGAN-GP for generating data to improve zero-day attack detection in IDS models. Firstly, the SA mechanism was incorporated to enhance the WGAN-GP's ability to understand long-range dependencies in data, enabling Generator G to create more coherent and detailed samples that better capture complex network traffic patterns. Secondly, an additional Discriminator D is introduced, combining the JS divergence with the Wasserstein distance. This design provides more subtle feedback to G, thereby improving discrimination and enhancing sample authenticity. Finally, the SA mechanism with the additional D was combined to form a cohesive model, the SA-JS-WGAN-GP, that maximizes these two advances to generate data similar to real network traffic. Extensive evaluation shows the proposed WGAN-GP models, particularly the SA-JS-WGAN-GP, are superior to the state-of-the-art GAN models in improving the performance of IDS detecting known and unknown network intrusions.

Future research will address overfitting and insufficient generalization in the SA-JS-WGAN-GP model trained on limited data through the use of transfer and ensemble learning techniques.  Moreover, the model will be evaluated on other benchmark datasets (CICIDS2017 and UNSW-NB15).

\section*{Acknowledgments}
This work was supported by the Engineering and Physical Sciences Research Council (EPSRC), UK (EP/W00366X/1).






\bibliographystyle{elsarticle-num} 
\bibliography{references}



\end{document}